\documentclass[prd,twocolumn,showpacs,preprintnumbers,amsmath,amssymb,nofootinbib]{revtex4}
\usepackage{graphicx}
\usepackage{psfrag}

\begin{document}
\title{Holographic melting and related properties of mesons in a quark gluon plasma}
\author{Kasper Peeters}
\email[Email:]{kasper.peeters@aei.mpg.de}
\affiliation{MPI f\"ur Gravitationsphysik,
  Am M\"uhlenberg 1,
  14476 Golm, Germany}
\author{Jacob Sonnenschein}
\email[Email:]{cobi@post.tau.ac.il}
\affiliation{School of Physics and Astronomy, The Raymond and Beverly Sackler Faculty of Exact Sciences,
Tel Aviv University, Ramat Aviv, 69978, Israel}
\author{Marija Zamaklar}
\email[Email:]{marija.zamaklar@aei.mpg.de}
\affiliation{MPI f\"ur Gravitationsphysik,
  Am M\"uhlenberg 1,
  14476 Golm, Germany}

\preprint{hep-th/0606195, AEI-2006-042, TAUP-2826/06}
\begin{abstract}
  We analyse mesons at finite temperature in a chiral,
  confining string dual. The temperature dependence of low-spin as
  well as high-spin meson masses is shown to exhibit a pattern
  familiar from the lattice. Furthermore, we find the dissociation
  temperature of mesons as a function of their spin, showing that at a
  fixed quark mass, mesons with larger spins dissociate at lower
  temperatures. The Goldstone bosons associated with chiral symmetry
  breaking are shown to disappear above the chiral symmetry
  restoration temperature. Finally, we show that holographic
  consideration imply that large-spin mesons do not experience drag
  effects when moving through the quark gluon plasma. They do,
  however, have a maximum velocity for fixed spin, beyond which they
  dissociate.
\end{abstract}

\maketitle

\section{Introduction and summary}

Of the various gravity duals to confining, non-supersymmetric gauge
theories in four dimensions, the Sakai-Sugimoto
model~\cite{Sakai:2004cn,Sakai:2005yt} stands out because it contains
chiral fermions as well as an associated chiral symmetry breaking
transition. This model has been analysed in quite some detail at zero
temperature. More recent work has analysed the phase diagram of the
model, showing an interesting structure in which the
confinement/deconfinement transition does not necessarily coincide
with the chiral symmetry breaking transition~\cite{Aharony:2006da}
(see also~\cite{Parnachev:2006dn}). In the present paper, we will
analyse the various phases which were found in~\cite{Aharony:2006da}
in more detail, focussing on the properties of mesons.

\begin{figure*}[t]
\psfrag{x4}{$x^4$}
\psfrag{t}{$t$}
\psfrag{Ul}{$u_\Lambda$}
\psfrag{Ub}{$u_0$}
\psfrag{Ut}{$u_T$}
\psfrag{U}{$u$}
\psfrag{pR}{$\pi R$}
\psfrag{B2}{$\beta/2$}
\psfrag{L}{$L$}
\includegraphics[width=.9\textwidth]{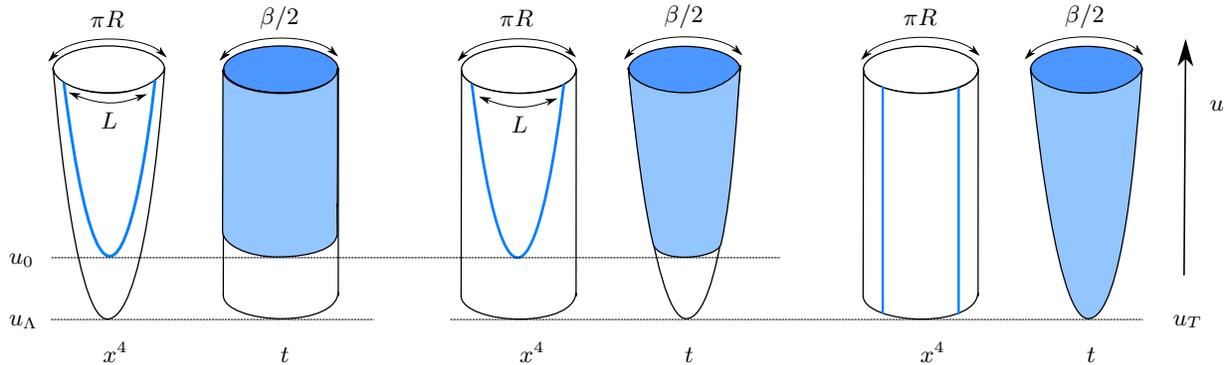}
\caption{The Sakai \& Sugimoto model in the low-temperature (left
  panel), intermediate-temperature (middle panel) and high-temperature
  (right panel) phases. Depicted are the two compact Euclidean
  directions~$x^4$ (the Kaluza-Klein circle) and~$t$ (Wick rotated
  time). The D8 embedding is shown as well (in blue).\label{f:threephases}}
\end{figure*}

The D4-brane background of the Sakai-Sugimoto model with Euclidean
signature exhibits two circle directions. The analysis
of~\cite{Aharony:2006da} has shown that this geometrical structure,
plus the presence of the D8-probes, leads to three different phases
(when the ratio~$L/R$, see figure~\ref{f:threephases}, is sufficiently
small). In the \emph{low-temperature} phase the background is the
Euclidean continuation of a Lorentzian background without any
horizon. Gluons are confined in this phase. After the
confinement/deconfinement transition for the gluons, we enter the
\emph{intermediate-temperature} phase, in which the Euclidean geometry
is instead given by the continuation of a black brane geometry. In
this phase gluons are deconfined, but chiral symmetry is still
broken. Mesonic bound states still exist, as the D8-brane embedding is
not yet touching the horizon. At sufficiently high temperature, the
lowest-energy configuration of the D8-branes is one in which they hang
vertically from infinity down to the horizon. This is the
\emph{high-temperature} phase, in which chiral symmetry is
restored. If the ratio~$L/R > 0.97$, there is no
intermediate-temperature phase, so that the confinement/deconfinement
and the chiral symmetry breaking transition coincide. For more
details, we refer to~\cite{Aharony:2006da}.

In the present paper, we will have a close look at the spectrum of
low-spin as well as high-spin mesons in these three phases. From both
lattice results and experiment, it is known that the meson spectrum
shows interesting behaviour as the temperature goes up. There is
abundant evidence, for instance, that heavy-quark bound states can
survive in a quark-gluon plasma to temperatures which are
substantially higher than the confinement/deconfinement
transition~\cite{deForcrand:2000jx}. Moreover, heavy-quark mesons tend
to survive to higher temperatures than light-quark mesons (see
e.g.~\cite{Karsch:2005nk,Wong:2006dz}). We would like to see to which
extent the Sakai-Sugimoto model is able to say something qualitatively
about such situations.
\medskip

Without further ado, let us briefly summarise the outcome of our
analysis. Firstly, we find that the temperature dependence of
\emph{both} small-spin (figure~\ref{f:mass_vs_T}) and large-spin
(figure~\ref{f:running-large}) mesons is qualitatively similar to what
is known from mesons in a quark-gluon
plasma~\mbox{\cite{Karsch:1999vy,Shuryak:2003xe,Kalinovsky:2005kx}}. That
is, the masses of mesons with low excitation number go down as the
temperature goes up.  The~$\rho$ and~$a_1$ meson masses approach each
other, although they do not become degenerate at the temperature where
chiral symmetry is restored.  For higher excitation level, the
temperature dependence appears to be more complicated
(figure~\ref{f:running}).

Secondly, we analyse the temperature dependence of large-spin mesons,
which are modelled by rotating U-shaped strings hanging from the probe
brane.  We find that for large-spin mesons there is a maximum value of
angular momentum beyond which mesons cannot exist and have to
``melt'', i.e.~dissociate (figure~\ref{spinmax}). For a given spin,
the maximum temperature can be substantially lower than the
temperature of the chiral symmetry restoration transition. We consider
this maximum-spin effect a prediction of the string/gauge-based
models, as it is quite generic and qualitatively independent of the
specifics of the background.

Thirdly, we find that large-spin mesons do not experience a drag
force. Instead, there is a critical velocity beyond which a fixed-spin
meson will dissociate. Equivalently, the maximum interquark distance
goes down as the velocity is increased (figure~\ref{f:LatJmax}).  The
absence of drag is unlike the situation for quark probes in a
quark-gluon
plasma~\cite{Herzog:2006gh,Casalderrey-Solana:2006rq,Gubser:2006bz,Herzog:2006se,Caceres:2006dj,Friess:2006aw,Sin:2006yz}.
For large-spin mesons at finite temperature, one can find generalised
solutions where the meson moves with linear velocity, rigidly, with
\emph{free} boundary conditions in the direction of motion
(figure~\ref{f:vy}). Hence one does not need to apply any force to
maintain this motion. In the dual language, this is a reflection of
the fact that if the quark gluon plasma is not hot enough to
dissociate mesons, then these colour singlets will not experience a
drag force generated by monopole interactions with the
medium.\footnote{The interaction with the medium could be still via
dipole (or higher moment) interactions. These, however, require
coupling of the string sigma model to the fields on the worldvolume of
the brane, and will be discussed elsewhere.}

\section{Review of the system at various temperatures}

In this section we review the various phases of the Sakai-Sugimoto
model, originally proposed in~\cite{Sakai:2004cn,Sakai:2005yt} and
analysed at finite-temperature in~\cite{Aharony:2006da}. Let us first
recall the Euclideanised model at zero temperature. The pure glue
theory is described by a background with a non-trivial metric, 
\begin{multline}
\label{D4}
{\rm d}s_{\text{con}}^2 = \left(\frac{u}{R_{\text{D4}}} \right)^{3/2}
  \big[ {\rm d}t^2 + \delta_{ij}\,{\rm d}x^i{\rm d}x^j  + f_\Lambda(u)
  ({\rm d} x^4)^2 \big] \\[1ex]
+  \left(\frac{R_{\text{D4}}}{u}\right )^{3/2}\left [\frac{{\rm d}u^2
  }{f_\Lambda(u)} + u^2 {\rm d}\Omega_4 \right ]\,,
\end{multline}
as well as a  non-constant dilaton and a four-form RR field strength
given by
\begin{equation}
\begin{aligned}
e^\phi &= g_s \left(\frac{u}{R_{\text{D4}}} \right)^{3/4}\,, \quad F_4=\frac
  {2\pi N_c}{V_4}\epsilon_4 \,,\\[1ex]
R_{\text{D4}}^3 &= \pi g_s N_c l_s^3\,,
\quad  f_\Lambda(u) := 1 - \left(\frac{u_\Lambda}{u}\right)^3 \,.
\end{aligned}
\end{equation}
Here~$u$ is the radial direction, which is bounded from below
by~$u\geq u_\Lambda$.  Our four-dimensional world is along the~$t$ and
$x^{i}$ $(i=1,2,3)$ directions. The main characteristic feature of the
background is the submanifold spanned by~$x^4$ and $u$, which has the
geometry of a cigar. The tip of the cigar is non-singular if and only
if the periodicity of~$x^4$ is
\begin{equation}
\delta x^4 =  \frac{4\pi}{3} \left( \frac{R_{\text{D4}}^3}{u_\Lambda}
\right)^{1/2} =: 2\pi R\, .
\end{equation}
Therefore, the circle~$x^4$ shrinks from its asymptotic size~$2\pi R$
down to zero and smoothly caps off at~$u=u_\Lambda$.  Chiral quarks in
the fundamental representation are incorporated by introducing a probe
D8-brane into the system. The probe fills out the full space, except
in the submanifold of the cigar, where it has a non-trivial profile.
At zero temperature, the spectrum of the glueballs (obtained from the
fluctuations of the supergravity fields of the background), as well as
of low-spin mesons (obtained from the fluctuations of the fields on
the probe brane) is \emph{discrete}, signalling that we are in the
confining phase.

Turning on the temperature~$T$ corresponds to the compactification of
the Euclidean time direction~$t$. The temperature is related to the
size~$\beta$ of the compact time direction through the standard field
theory relation $\beta=1/T$. We now have two circle directions, one
being the asymptotic circle of the cigar discussed above, with
period~$2\pi R$, and the other one being the thermal circle, with
period~$\beta$. This geometry is depicted in the left panel of
figure~\ref{f:threephases}, which also shows the embedding of the
D8-brane.  The glueball and meson spectra in this phase are still
discrete. In fact, they are unmodified with respect to the zero
temperature situation, given that the metric is unmodified (modulo the
global identification). We will refer to this phase as the
\emph{low-temperature} phase.

As the temperature of the system is increased, the system will at a
certain point undergo a first order phase transition in which the
\emph{gluonic} degrees of freedom get deconfined. This is reflected in
the change of the background metric from~\eqref{D4} to
\begin{multline}
\label{e:intermediateTmetric}
{\rm d}s_{\text{decon}}^2 =
\left(\frac{u}{R_{\text{D4}}}\right)^{3/2} \big[ f_T(u){\rm d}t^2 + \delta_{ij} {\rm d}x^i
{\rm d}x^j + ({\rm d}x^4)^2\big] \\[1ex]
+\left(\frac{R_{\text{D4}}}{u}\right)^{3/2} \left[\frac{{\rm d}u^2}{f_T(u)} + u^2{\rm
  d}\Omega_4^2\right]\,, 
\end{multline}
\begin{equation*}
\text{with}\quad f_T(u) := 1 - \left(\frac{u_T}{u}\right)^3  \, .
\end{equation*}
The metric has the same form as the low-temperature metric~\eqref{D4},
but with the role of the~$t$ and~$x^4$ directions exchanged. This is
depicted in the middle and right panels of
figure~\ref{f:threephases}. The phase transition between the
metric~\eqref{D4} and the metric~\eqref{e:intermediateTmetric} happens
when their free energies are equal, which is when the two circles have
the same radius, i.e.~when $T = T_c = 1/2\pi R$. This transition is
first-order because the solutions do not smoothly connect, and
continue to exist as separate solutions both below and above the
transition. The spectrum of the glueball fluctuations in the phase
described by~\eqref{e:intermediateTmetric} is \emph{continuous},
signalling the deconfinement of the gluonic degrees of freedom.

An order parameter of this phase transition is the Polyakov loop,
corresponding to a string wrapped around the time direction. In the
low-temperature phase the time circle is non-contractible and not the
boundary of a disc, so that the Polyakov loop vanishes. After the
transition, this circle becomes contractible, resulting in a non-zero
expectation value for the Polyakov loop. Other order parameters are
the Wilson loop (which has a linear quark/anti-quark potential in the
first background but vanishing tension in the second) and the
behaviour of the free energy as a function of~$N_c$ (namely~$\sim
N_c^0$ for the first background and~$\sim N_c^2$ for the second).

In contrast to the gluonic degrees of freedom, the matter degrees of
freedom may still have bound states in this phase. In order to analyse the
matter phases, one needs to determine the shape of the D8-brane, which
can be obtained from the DBI part of the effective action,
\begin{equation}
S_{\text{DBI}} = T_8\,\int\!{\rm d}t{\rm d}^3x{\rm d}x^4 {\rm d}\Omega_4\,
e^{-\phi} \sqrt{-\det \hat{g}}\,.
\end{equation}
In the two backgrounds~\eqref{D4} and~\eqref{e:intermediateTmetric},
the DBI actions have the following form,
\begin{equation}
\begin{aligned}
S^{\text{con}}_{\text{DBI}} &= \frac{\hat T_8 }{g_s}\int\!{\rm d}x^4\, u^4 \sqrt{f(u)+ 
\frac{R^3_{D4}}{u^3} \frac{{u'}^2}{f(u)}}\,,\\[1ex]
S^{\text{decon}}_{\text{DBI}} &=  \frac{\hat T_8 }{g_s}\int\!{\rm d}x^4\, u^4 \sqrt{f(u)+ 
\frac{R_{D4}^3}{u^3} {u'}^2}\,,
\end{aligned}
\end{equation}
where from now on we will drop the subscript on~$f$ as it is clear
from the context which one should be used.  In the confining
background, the solution to the equations of motion (depicted by the
blue curve on the left panel of figure~\ref{f:threephases}) is given by
\begin{multline}
\label{t2}
x^4(u) = \\[1ex]
u_0^4\, f(u_0)^{1/2} \int_{u_0}^u\frac{{\rm
	 d}u}{\displaystyle \left(\frac{u}{R_{\text{D4}}}\right)^{3/2}
f(u)\sqrt{u^8 f(u) - u_0^8 f(u_0)}}\,.
\end{multline}
The D8-brane solution in the middle panel is related to this one by the simple expression
\begin{equation}
\label{t1}
\frac{{\rm d}x^4_{\text{decon}}}{{\rm d}u} = \sqrt{f(u)}
\frac{\, \, {\rm d} x^4_{\text{con}}}{{\rm d}u}\,.
\end{equation}

However, there is a second configuration in the deconfined
background, in which the embedding is simply given
by~$x^4=\text{const}.$ (depicted in the third panel of
figure~\ref{f:threephases}). The most important fact about this third
phase is that chiral symmetry is restored. Because there are two
stacks of branes, every mode will appear twice, as a representation
of~$\text{U}(N_f)_\text{L}$ and of~$\text{U}(N_f)_\text{R}$.  So we
potentially have two phases in the deconfined background, which we
will call the \emph{intermediate}- and \emph{high}-temperature
phase. Whether or not the intermediate-temperature phase is realised
depends on the ratio~$L/R$, where~$L$ is the asymptotic distance
between the D8 and anti-D8 (see
figure~\ref{f:threephases}). In~\cite{Aharony:2006da} it was found
that for~$L/R>0.97$, the intermediate phase is absent altogether,
i.e.~the confinement/deconfinement and chiral symmetry restoration
transitions occur simultaneously. One can map the parameter~$L/R$ to a
gauge dynamical parameter by noting that~$1/R$ is the scale of the
glueball masses, while (as we will show in section~\ref{s:lowspin}),
$1/L$~is proportional to the mass of mesons composed of quarks of
large constituent mass, $u_0 \gg u_T$. We refer the reader
to~\cite{Aharony:2006da} for more details on the full phase diagram,
including the region where the supergravity picture is not valid.

In the string/gauge duality models, one associates low-spin mesons
to fluctuations of the gauge fields and (pseudo) scalars that live on
the probe branes, while high-spin ones are associated to spinning string
configurations. We will address the thermal spectrum of the former in
section~\ref{s:lowspin} and the latter in~\ref{s:highspin}.

\section{Low-spin mesons at intermediate temperature}
\label{s:lowspin}

In this section we will analyse the spectrum of low-spin mesons (with
spin~$\leq 1$) in the intermediate-temperature regime. The spectrum of
low spin mesons in the \emph{low-temperature} phase is unmodified with
respect to zero temperature.  This is a consequence of the fact that
the (Euclidean) metric is globally unmodified (the only difference is
that there is now a global periodic identification of the Euclidean
time direction). This perhaps unexpected feature seems to be a generic
property of large-$N_c$
theories~\cite{Neri:1983ic,Pisarski:1983db,Aharony:2006da} (see also
section~\ref{s:low2interm}).

In the intermediate-temperature phase we expect that the spectrum of
low-spin mesons is discrete, because the probe does not intersect the
horizon. This is similar to the mechanism which ensures discreteness
of the spectrum at zero temperature. Furthermore, given that the
effective tension of strings near the brane decreases with the
increase of temperature, one expects that the masses of the mesons
decrease as the probe brane comes closer to the horizon, or
equivalently, as the temperature is increased. Indeed, our explicit
computation shows this behaviour.

Our starting point is the metric of the background describing the hot
gluonic plasma, as given in~\eqref{e:intermediateTmetric}. This background leads to
an induced metric on the D8-brane world-volume, which reads
\begin{multline}
{\rm d}\hat{s}_{\text{interm}}^2 = 
\left(\frac{u}{R_{\text{D4}}}\right)^{3/2} \left(f(u){\rm d}t^2 + \delta_{ij} {\rm d}x^i
{\rm d}x^j\right) \\[1ex]
+ \left[ \left(\frac{R_{\text{D4}}}{u}\right)^{3/2} \frac{1}{f(u)} 
         + \left(\frac{{\rm d}x^4}{{\rm d}u}\right)^2\, \left(\frac{u}{R_{\text{D4}}}\right)^{3/2} \right]
  {\rm d}u^2\\[1ex]
+ \left(\frac{R_{\text{D4}}}{u}\right)^{3/2} u^2 {\rm d}\Omega_4^2\,.
\end{multline}
We are interested in computing the spectrum of \emph{vector} mesons,
by considering small fluctuations on the worldvolume gauge fields of
the probe D8-brane (scalar and pseudoscalar mesons can be treated with
similar methods but will not be discussed here).  In order to do this,
we first expand the gauge field on the worldvolume of the D8-brane as~\cite{Sakai:2004cn}
\begin{equation}
\label{Fe}
\begin{aligned}
F_{\mu\nu} &= \sum_{n} F_{\mu\nu}^{(n)}(x^\rho)\,\psi_n(u)\,,\\[1ex]
F_{\mu u}  &= \sum_{n} \partial_\mu\varphi^{(n)}\,\phi_n(u) -
B^{(n)}_\mu \partial_u\psi_n(u) \\[1ex]
&= \partial_\mu \varphi^{(0)}\,\phi_{0}
  + \sum_{n\ge 1} \left( \partial_\mu \varphi^{(n)} - B^{(n)}_\mu \right)\partial_u \psi_{(n)}\,.
\end{aligned}
\end{equation}
where the last line is obtained by taking $\phi_{(n)} = m_n^{-1}
\partial_u \psi_{(n)}(u)$.  To simplify the consideration, we choose
to focus on the space-like components~$B_i$ of the vector
fields. Moreover, we only consider the masses defined by
\begin{equation}
\label{e:massdef}
\partial_0^2 B^{(n)}_i = - m_n^2 B^{(n)}_i \, , 
\end{equation}
that is, we consider the behaviour of the pole mass rather than the
screening mass. Thus, we consider only spatially homogeneous modes,
i.e.~we consider the equation of motion for fields
satisfying~$\partial_i B^{(n)}_j = 0$. \footnote{For non-homogeneous modes
the definition of mass becomes more subtle.  However, since we expect
that the homogeneous modes considered here are rich enough to
demonstrate all relevant features of the spectrum as a function of the
temperature, we restrict our attention to this case.}  In this case,
after using the equations explicit probe brane embedding~\eqref{t1}
and~\eqref{t2} the probe brane action reduces to
\begin{figure}[t]
\begin{center}
\psfrag{Lsu0}{$L u_0^{1/2}$}
\psfrag{yT}{$y_T$}
\includegraphics[width=.9\columnwidth]{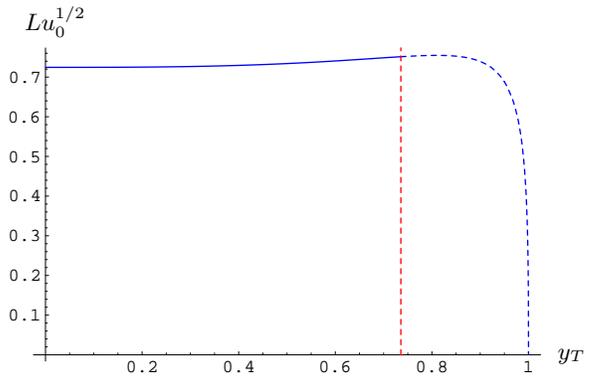}
\end{center}
\caption{The asymptotic distance~$L$ between the D8 and anti-D8 as a
  function of~\mbox{$y_T := u_T/u_0$}. The red dashed line indicates
  the phase transition to the high-temperature phase, above which one
  should instead use two parallel D8 stacks. Note that ``$L =
  \text{const.} \rightarrow u_0 = \text{const.}$''  holds to good
  approximation in the intermediate-temperature regime.\label{f:L_vs_yT}}
\end{figure}%
\begin{figure*}[t]
\begin{center}
\psfrag{m2}{$m^2$}
\psfrag{n}{$n$}
\psfrag{Ceven}{C-even}
\psfrag{Codd}{C-odd}
\includegraphics[width=.47\textwidth]{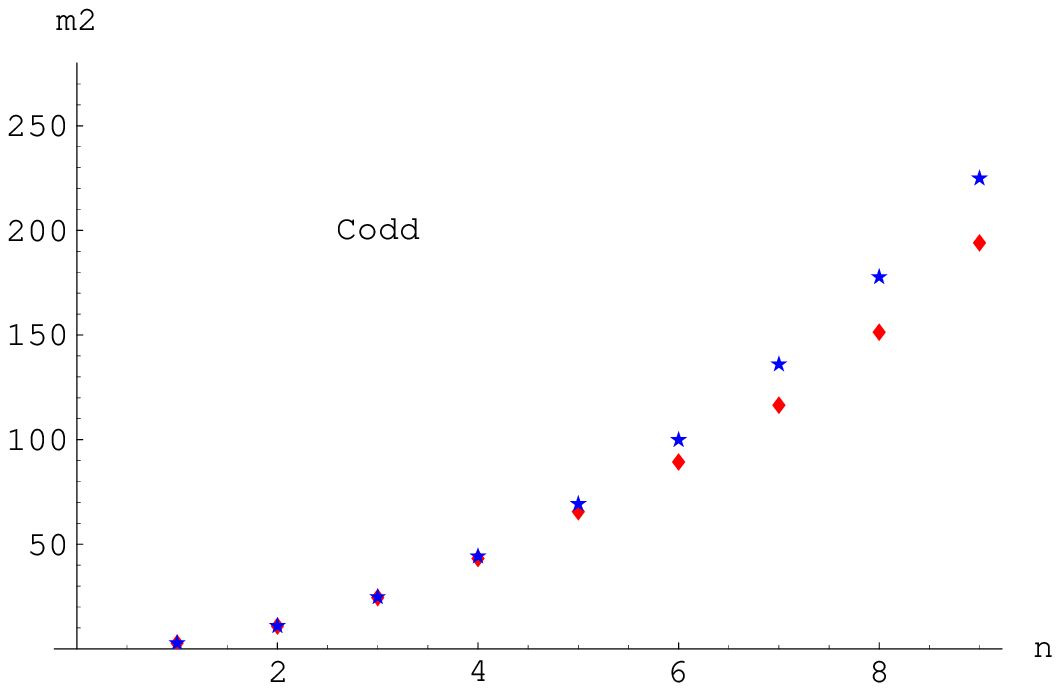}\qquad
\includegraphics[width=.47\textwidth]{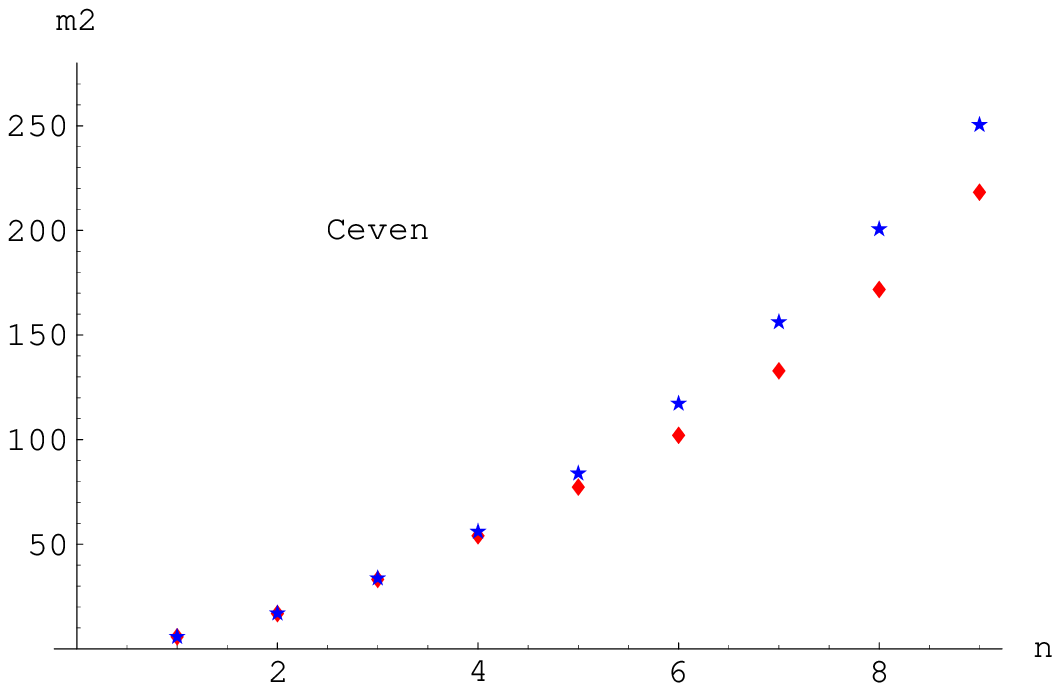}
\end{center}
\caption{Meson masses (squared) as a function of level~$n$ for the
  zero temperature (upper dots) and intermediate-temperature phases
  (lower dots), for an asymptotic separation between the D8 and
  anti-D8 fixed at~$L/R=0.63$, with~$u_\Lambda=u_T=1$.
  (Extensions of these curves to even higher level show~$m^2
  \sim n^2$ behaviour for large~$n$, both at zero and at finite
  temperature, which is one of the unrealistic features of these
  models~\cite{Schreiber:2004ie,Shifman:2005zn,Karch:2006pv}).\label{f:masses_finite_T}}
\end{figure*}%
\begin{multline}
\hat{S}_{\text{trunc}} = 
  \int\!{\rm d}^4x {\rm d}u\;
  u^4 \gamma^{1/2}
  \left(\frac{R_{\text{D4}}}{u}\right)^{3/2}\,f(u)^{1/2}\\[1ex]
\times 
\begin{aligned}[t]
& \bigg[ \frac{1}{\gamma\,f(u)} (\partial_0\varphi^{(0)})^2\,\phi^{(0)}\phi^{(0)}\\[1ex]
&-\frac{1}{f(u)}\left(\frac{R_{\text{D4}}}{u}\right)^3\, 
    \partial_0 B^{(m)}_i \partial_0 B_{(n)}^i\,\psi_{(m)}\psi_{(n)}\\[1ex]
&+ \gamma^{-1} B^{(m)}_i B_{(n)}^{i}\,\partial_u \psi_{(m)} \partial_u
  \psi_{(n)}\bigg]\,, 
\end{aligned}
\end{multline}
\begin{equation*}
\text{with}\qquad \gamma \equiv  {\frac{u^8}{u^8 f(u) - u_0^8 f(u_0)}} \, .
\end{equation*}
After a partial integration with respect to the $u$-coordinate, the
equation of motion for the field~$B_i^{(m)}$ becomes
\begin{multline}
\frac{ u^4 \gamma^{1/2} }{f(u)^{1/2}} \left(\frac{R_{\text{D4}}}{u}\right)^{9/2} \partial_0^2 B^{(n)}_i \psi_{(n)}\\[1ex]
- \partial_u\left( u^4 \gamma^{-1/2} f(u)^{1/2}\, \left(\frac{R_{\text{D4}}}{u}\right)^{3/2}\, \partial_u \psi_{(n)} \right) B^{(n)}_i = 0\, \, .
\end{multline}
This equation will reduce to the canonical form (with thermal
masses~$m_n$ as in~\eqref{e:massdef}) if the modes $\psi_{(n)}$
satisfy the equation
\begin{multline}
\label{midT}
-u^{1/2} \gamma^{-1/2}\,f(u)^{1/2}\,\partial_u\left( u^{5/2}
 \gamma^{-1/2} f(u)^{1/2}\partial_u \psi_{(n)}\right) \\[1ex]
= R_{\text{D4}}^3 \, m_n^2\, \psi_{(n)}\,.
\end{multline}
Equation~\eqref{midT} is very similar to the equation in the zero
temperature case.  The only difference is the appearance of extra
factors~$f(u)^{1/2}$ in the term on the left-hand side.  In order to
have canonically normalised kinetic terms, the modes should also
satisfy the normalisation conditions
\begin{equation}
\begin{aligned}
\int_{u_0}^\infty \!{\rm d}u\, u^4 \gamma^{1/2} f^{-1/2}(u)
\left(\frac{u}{R_{\text{D4}}}\right)^{9/2}\,\psi_{(m)}\psi_{(n)} &=
\delta_{mn}\,,\\[1ex]
\int_{u_0}^\infty \!{\rm d}u\, u^4 \gamma^{-1/2} f^{-1/2}(u)
\left(\frac{R_{\text{D4}}}{u}\right)^{3/2}\, \phi^{(0)} \phi^{(0)} &= 1\,.
\end{aligned}
\end{equation}
The zero mode~$\phi^{(0)} = u^{-5/2} f(u)^{-1/2} \gamma^{1/2}$ is
normalisable with this norm (there is no problem at the horizon
because~$u_0 > u_T$), and so we see that there is still a massless
pion present in the intermediate-temperature phase.

Before applying numerical methods to compute the spectrum, it is
interesting to note that in the limit of $u_0 \gg u_T$ the spectrum
simplifies and one can easily determine the scale of the meson
masses.  In this limit, which corresponds to a small separation distance
between the stacks of branes and anti-branes $L \ll R$, the thermal
factor $f(u)\rightarrow 1$ and in particular also $f(u_0)\rightarrow
1$. Therefore,
\begin{equation}
\gamma\equiv\frac{u^8}{u^8 f(u) -
 u_0^8 f(u_0)}\rightarrow \frac{1}{1-y^{-8}}
\end{equation}
where the dimensionless quantity $y\equiv u/u_0$.
In fact, for this case we can rewrite~\eqref{midT} in terms of~$y$
in the following form
\begin{equation}
- y^{1/2} \gamma^{-1/2}(y) \partial_y \left(
   y^{5/2} \gamma^{-1/2}(y) \partial_y \psi_{(n)}
\right) = \frac{R_{\text{D4}}^{3}}{u_0}\,m_{n}^2\,\psi_{(n)} \,.
\end{equation}
Now since the left-hand side is expressed in terms of the
dimensionless quantity~$y$, the right-hand side should also be
dimensionless which implies that
\begin{equation}
\label{scaling}
m_n^2 \sim \frac{u_0}{R_{\text{D4}}^3}
\end{equation}
Now since $u_0\sim 1/L^2$ (see below), the final conclusion is that
the mass of these ``short'' mesons scales as
\begin{equation}
M_{\text{meson}}\sim \frac{1}{L}\,.
\end{equation}

\begin{figure}[t]
\begin{center}
\psfrag{rho}{``$\rho$''}
\psfrag{a1}{``$a_1$''}
\psfrag{m2}{$m^2$}
\psfrag{T}{$T/T_c$}
\includegraphics[width=.9\columnwidth]{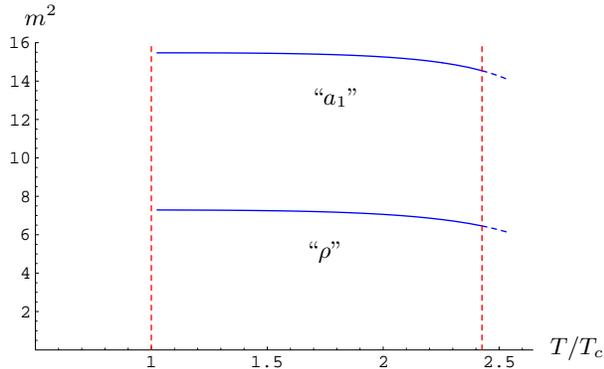}
\end{center}
\caption{Squared masses of the two lightest vector mesons as a
  function of temperature in the intermediate-temperature regime,
  for~$L=0.256 < 0.97\, R$. Masses are normalised as in~\protect\eqref{midT}.
 The decrease of the mass is in qualitative agreement with results
  from other models and experiments~\cite{Shuryak:2003xe}, though one
  should be cautious since the latter are not in the large-$N_c$
  limit. The $\rho$ and $a_1$ do not become degenerate in our
  model.\label{f:mass_vs_T}}
\end{figure}

To find the remainder of the spectrum, we can solve
equation~\eqref{midT} numerically using a shooting technique. We will
keep the asymptotic separation~$L$ between the D8 and anti-D8 stacks
fixed, and use the expression
\begin{equation}
L = \int\!{\rm d}x^4 = \int_{u_0}^\infty\!\frac{{\rm d}u}{u'}
\end{equation}
to determine~$u_0$ for a given temperature (see also
figure~\ref{f:L_vs_yT}).  The results for the thermal masses~$m^2_n$
of the vector mesons are depicted in
figure~\ref{f:masses_finite_T}. The spectrum is discrete, labelled by
the mode number~$n$ and parity. By comparing with the zero-temperature
result, we observe that the masses of light mesons decrease as the
temperature is increased. The temperature dependence of the masses of
the ``$\rho$'' and ``$a_1$'' mesons are shown in
figure~\ref{f:mass_vs_T}. 

Our findings are in qualitative agreement with lattice
computations~\cite{Karsch:1999vy,Gottlieb:1996ae} and the results of
the ``phenomenological'' model of Karch et al.~\cite{Ghoroku:2005kg},
although we do not see the degeneracy of the~$\rho$ and~$a_1$ meson
masses at high temperature. Moreover, we will see in
section~\ref{s:highT} that the spectrum becomes continuous in the
phase where chiral symmetry is restored. Therefore, the phase
transition from intermediate to high temperature, which is first-order
in our model, is rather different from QCD. However, what we see is
consistent with the fact that mesons in the chirally symmetric phase
are extremely unstable in the large-$N_c$ limit: their decay widths
scale as a positive power of~$N_c$.\footnote{For results of other,
non-holographic models see e.g.~the summary
in~\cite{Bhattacharyya:1995xt}.}

This qualitative behaviour of the spectrum in our model (and the
ad-hoc model of~\cite{Ghoroku:2005kg}) is a direct consequence of the
fact that the flavour branes are, for fixed~$L$, closer to the horizon
when the temperature is higher. Hence, this decrease of masses with
temperature will be a common feature for all gravitational backgrounds
which contain a horizon. However, the precise behaviour of the masses
as a function of temperature is clearly model dependent.  In our case,
it depends on parity, the level~$n$ and the spin of the meson, as well
as on the constituent quark masses. It is interesting to note that
generically, for fixed spin, level and parity, the meson mass runs
slower for larger constituent quark masses than for lighter quarks.
This behaviour is again in qualitative agreement with lattice
simulations, see for
example~\cite{Asakawa:2003re,Datta:2003ww,Wissel:2005pb,Iida:2005ea}
for quenched results and~\cite{Morrin:2005zq} for a confirmation in
unquenched QCD.  There, it is found that deeply bound heavy quark
states, such as the $\eta_c$ ($J=0$) or the $J/\Psi$ ($J=1$), survive
up to temperatures which are $\leq 2\, T_{\text{crit}}$, where
$T_{\text{crit}}$ is the temperature where the light mesons $\rho,
\omega$ and $\phi$ melt.  For our system, this behaviour is a natural
consequence of the fact that the constituent quark mass is, roughly
speaking, related to the distance of the tip of the probe brane to the
horizon.  If this distance is increased (while keeping the horizon
temperature fixed in units of the radius of the circle~$x^4$), a meson
of the same spin, parity and level will correspond to an excitation of
the brane which is further away from the horizon, and hence less
affected by the temperature. Note that because the phenomenological
model of~\cite{Ghoroku:2005kg} lacks probe branes, it does not have
this parameter at its disposal. Hence, it thus seems harder to
reproduce this feature of the mesonic spectrum in such ad-hoc models.

\begin{figure}[t]
\begin{center}
\vspace{2ex}
\psfrag{m2}{$m^2$}
\psfrag{T}{$\;T/T_c$}
\includegraphics[width=.9\columnwidth]{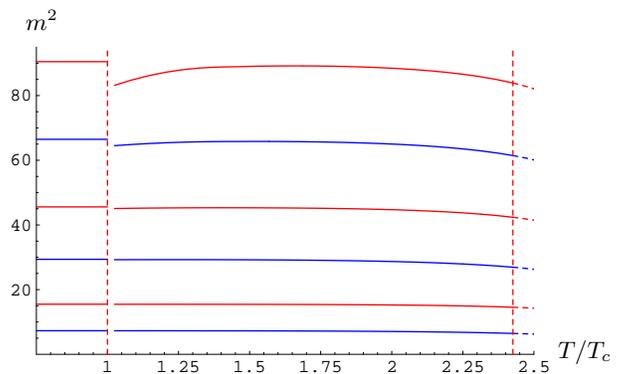}
\end{center}
\caption{Behaviour of six lowest-mass vector mesons (for fixed~$L/R$)
  as a function of temperature in the intermediate-temperature regime,
  and a comparison with the masses in the low-temperature phase. Note
  the jump in the spectrum at the confinement/deconfinement
  transition, and also note that the masses of the higher excited
  modes do not exhibit monotonic behaviour as a function of
  temperature.\label{f:running}}
\end{figure}

One can also follow the mass trajectories for the higher excited
modes. This results in figure~\ref{f:running}. This plot shows
interesting qualitative similarities with results in real-world QCD,
see e.g.~\cite{Shuryak:2003xe}, although the latter results are not in
the large-$N_c$ limit so one should be careful with a direct
comparison. It would be very interesting to extract more quantitative
dependence of the meson masses on the aforementioned set of parameters
from our model. We leave this question for future investigation.

\subsection{Phase transition from low to intermediate temperature}
\label{s:low2interm}

As mentioned before, the mesonic spectrum in the low-temperature phase
is unchanged as the temperature is increased. Although this may sound
counterintuitive, it seems to be a generic feature of large-$N_c$
theories~\cite{Neri:1983ic,Pisarski:1983db}. Namely, it was shown
in~\cite{Neri:1983ic,Pisarski:1983db} that, for example, the value of
the chiral condensate~$\langle \bar{\psi} \psi\rangle$ is
\emph{independent} of the temperature, essentially because in the
confining phase the theory behaves effectively as a gas of
non-interacting glueballs and mesons.  This behaviour is indeed what
is reproduced from a \emph{generic} supergravity solution, since the
metric for finite, low temperatures is the same as the metric for zero
temperature (note that the same remains true even if $N_f \sim N_c$
and one constructs the back-reacted geometry). To incorporate
finite-temperature effects on the meson spectrum and the value of the
chiral condensate, one would need to go beyond the supergravity
approximation and deal with genuine finite-temperature string
theory. 

Let us nevertheless see how the intermediate-temperature and
low-temperature phases are connected in our model. To do this, we send
the temperature to the critical value~$T_c$ where the gluonic degrees
of freedom get deconfined (i.e.~$T \rightarrow 1/2\pi
R$). Figure~\ref{f:running} shows that the mesonic spectrum at
intermediate temperature is \emph{not} connected to the spectrum in
the low-temperature regime. However, since as argued
in~\cite{Aharony:2006da} the transition is first-order, such a jump
should be expected.  Do notice, however, that the mesons are only
mildly affected by the ``violent'' transition of the gluonic degrees
of freedom. The mesonic spectrum, though discontinuous, is not
rearranged: for every meson state before the transition, there is a
corresponding state after the transition. In a certain sense the
mesons are merely spectators in this phase transition.

\section{High-spin mesons at intermediate temperature}
\label{s:highspin}

In this section we examine the properties of \emph{high-spin} mesons
as a function of the temperature. For low-spin mesons, we have seen
that the dependence of the string tension on the temperature
determines the behaviour of the masses. We expect this mechanism to
apply to high-spin mesons as well. Therefore we expect that for a
meson of fixed spin and low excitation number, the mass will decrease
as a function of temperature. We will indeed find this
behaviour. However, there is now an additional interesting phenomenon,
which only appears for high-spin mesons, since their spin can vary
significantly. Spin is a parameter which controls the length of the
string as well the distance of the tip of the string to the horizon.
Hence, a natural question arises as to whether the
confinement--deconfinement transition of mesons (i.e.~the process of
``dissociation'' of the mesons), takes place at the same temperature
for all mesons, irrespective of their masses and spins?  Intuitively,
one expects that higher-spin mesons will dissociate first, given that
they ``hang'' closer to the horizon. The question is then what is the
critical value of the spin as a function of temperature?

To answer the questions posed in the previous paragraph, let us
consider large spinning mesons at finite temperature. These are
modelled by rotating U-shape strings hanging from the probe brane in
the middle picture of figure~\ref{f:threephases}.  The relevant part
of the background metric~\eqref{e:intermediateTmetric} is
\begin{multline}
{\rm d}s^2 = \left(\frac{u}{R_{\text{D4}}}\right)^{3/2} \left( - f(u)\, {\rm d}t^2
+ {\rm d} \rho^2 + \rho^2\,{\rm d} \varphi^2 \right) \\[1ex]
+
\left(\frac{R_{\text{D4}}}{u}\right)^{3/2} \frac{{\rm d}u^2}{f(u)}\, .
\end{multline}
We go to the static gauge for the string action and make the following
ansatz for the rotating configuration,\footnote{When solving the
equations of motion numerically, the fact that this gauge leads to a
${\rm d}u/{\rm d}\sigma$-derivative which blows up at the flavour
brane is rather inconvenient. For numerical purposes, a better gauge
choice is, for instance, given by~$\rho + u = \sigma$.}
\begin{equation}
\label{ansatz-meson}
t = \tau \, , \quad \rho=\sigma\,,\quad u = u(\rho)\,,
 \quad \varphi = \omega \tau  \, .
\end{equation}
We see that the ansatz has the the same form as in the
zero-temperature case~\cite{Kruczenski:2004me}. Hence, just like for
the Wilson loop, the only effect of finite temperature will be that
the details of the shape~$u(\sigma)$ change as the temperature is
increased.

\begin{figure}[t]
\begin{center}
\psfrag{u}{$u$}
\psfrag{rho}{$\rho$}
\includegraphics[width=.9\columnwidth]{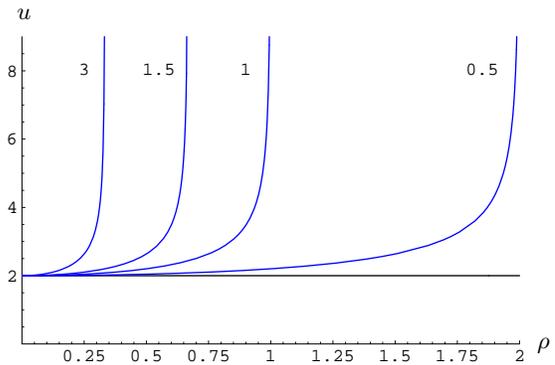}
\end{center}
\caption{The boundary curve~\protect\eqref{cond}. Rotating strings
  have to lie above this curve in order for their action to be
  real. The various curves correspond to various values of the
  frequency~$\omega$.
  The horizon is located at~$u_T=2$. These curves are independent of
  the probe brane location.\label{f:bound-rigid}}
\end{figure}

\begin{figure*}[t]
\begin{center}
\vspace{2ex}
\psfrag{u}{$u$}
\psfrag{rho}{$\rho$}
\includegraphics[width=.45\textwidth]{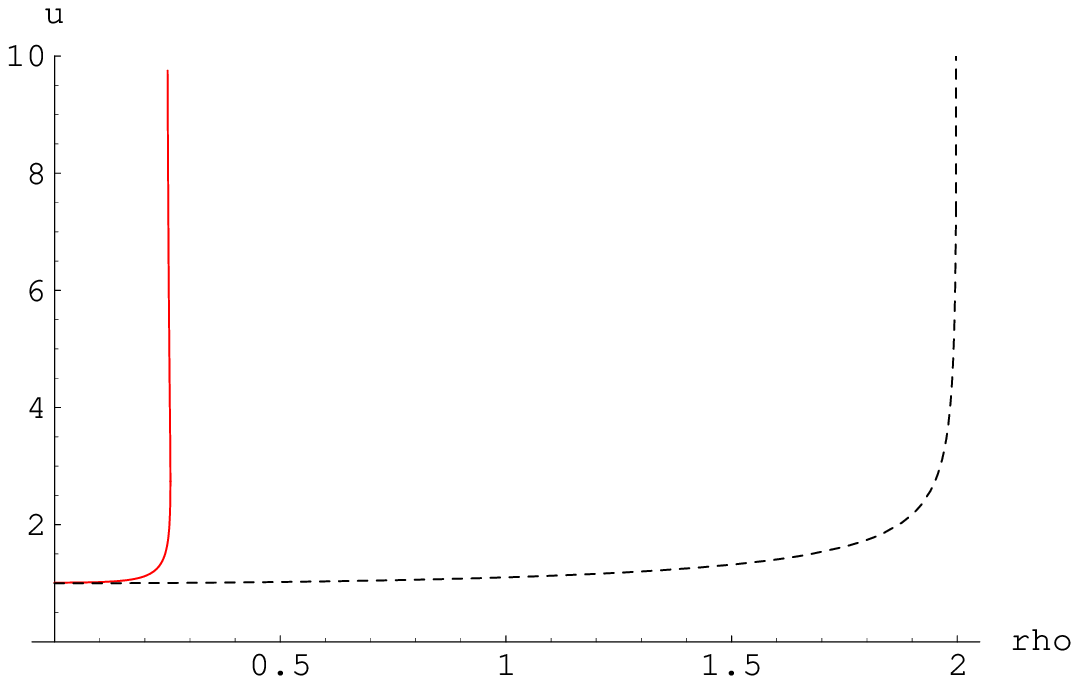}\qquad
\includegraphics[width=.45\textwidth]{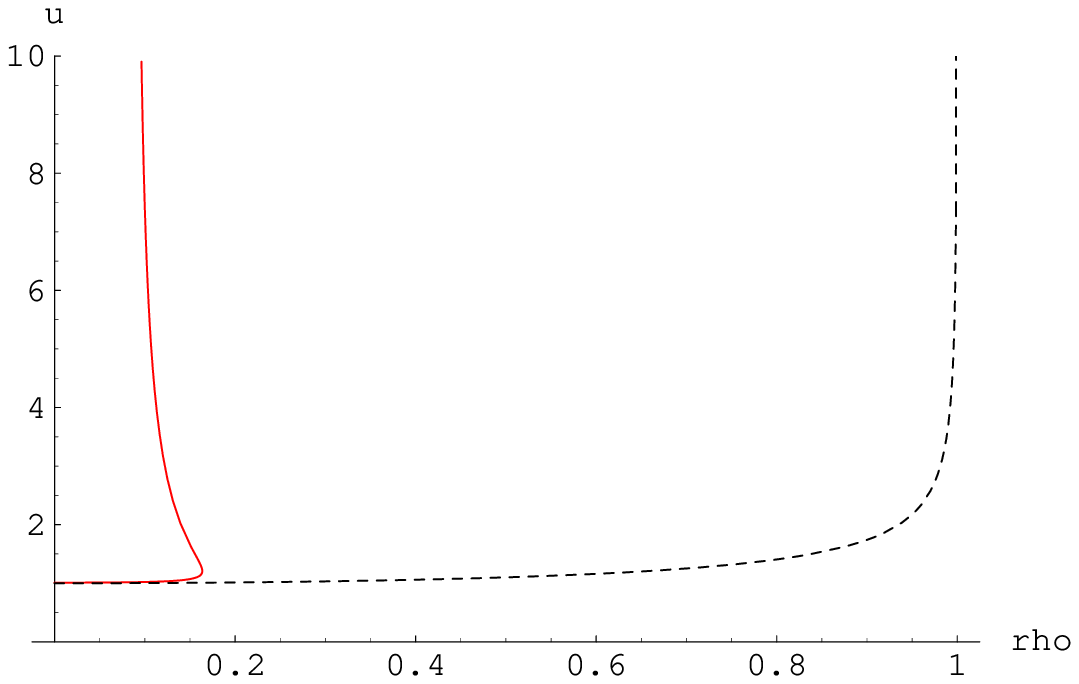}
\end{center}
\caption{Comparison of numerical solutions (solid curves) with the
  boundary curve~\protect\eqref{cond} (dashes), for~$\omega=0.5$
  and~$\omega=1$ respectively. Although the numerical solution lies
  manifestly above the boundary curve, as it should, it deviates quite a
  bit from it. The numerical solutions where obtained by
  fixing~$u(0)=u_T+\epsilon$ for a small~$\epsilon$, as well
  as~\mbox{$u'(0)=0$}.\label{f:bound}}
\end{figure*}%

One could wonder why such an unmodified ansatz is correct, given that
the string is moving with angular velocity, and velocity dependence
has been shown to modify the zero-temperature ansatz for vertical
strings due to ``dragging''
effects~\cite{Herzog:2006gh,Casalderrey-Solana:2006rq,Gubser:2006bz,Herzog:2006se,Caceres:2006dj,Friess:2006aw,Sin:2006yz}.  To answer
this question we need to see if the ansatz (\ref{ansatz-meson}) allows
for physical solutions. If it does, this would imply that
\emph{angular} velocity of a meson in a hot medium does not lead to the
distortion of rotation, i.e.  one does not need to apply a force to
maintain the rigid rotation: an initially rigidly rotating meson
continues to rotate rigidly. This of course does not mean that the
$u(\rho)$ profile is unaffected by the temperature. A separate
question is what is the effect of \emph{linear} motion of this colour
singlet. This will be explored in section~\ref{s:drag}.

To answer the first question, about the possibility of having a
rigidly rotating meson in a quark gluon plasma, let us compute the
action for the meson~\eqref{ansatz-meson}. This ansatz leads to
\begin{equation}
\label{Sind}
S = \int\!{\rm d}\tau \,  {\rm d}\rho \, \sqrt{\left(\frac{u}{R_{\text{D4}}}\right)^3\,\left(\rho'^2  + \frac{u'^2}{f(u)} \frac{R_{\text{D4}}^3}{u^3}\right) \left( f(u)  - \rho^2\omega^2 \right)} \,  .
\end{equation}
Before we analyse the resulting equations of motion, let us first make
a few general remarks. We see that positivity of the argument of
the square root in~\eqref{Sind} requires that \mbox{$f(u)> \rho^2
\omega^2$}. This in turn means that for a given angular frequency
$\omega$, the string solution $u(\rho)$ has to lie above the curve
\begin{equation} 
\label{cond}
u(\rho) \geq \frac{u_T}{(1 - \rho^2 \omega^2)^{1/3}} \, .
\end{equation}
In figure~\ref{f:bound-rigid} we have depicted these curves for
various values of $\omega$. Firstly, we see that the tip of any string
(the $\rho=0$ point) is in principle allowed to touch the horizon for
\emph{any} angular frequency~$\omega$.  Secondly, for given~$\omega$,
the maximal allowed extent of the string is determined by the
intersection of the curve with~$u_0$, and is given by
\begin{equation} 
\rho_{\text{max}} = \frac{1}{\omega}\sqrt{1 - \frac{u_T}{u_0}} \, .
\end{equation}
(One could have hoped that the boundary curve~\eqref{cond} becomes a
good approximation to the real string solution in some
regime. However, this does not seem to be the case, as is easily seen
from a comparison of this curve with some real solutions, figure~\ref{f:bound}.)

The equations of motion following from the action~\eqref{Sind} are
given by
\begin{multline}
\label{equ}
 -2 \sqrt{...} \frac{{\rm d}}{{\rm d}\sigma} \left( \frac{1}{\sqrt{...}} \frac{u'}{f(u)} (f(u) - \rho^2 \omega^2) \right)   \\[1ex]
+ f'(u)\bigg(\frac{u'^2 \rho^2 \omega^2}{f(u)^2} + \frac{u^3}{R_{\text{D4}}^3} 
(\rho')^2 \bigg) \\[1ex]
+ \frac{3 u^2}{R_{\text{D4}}^3} \bigg((\rho')^2  -(\rho')^2 \rho^2 \omega^2 \bigg) = 0\, ,
\end{multline}
where $\sqrt{...}$ is the density of Nambu-Goto
action~\eqref{Sind}. One can verify that varying the action first and
then inserting the ansatz yields the same result. In addition we have
to take care of the boundary terms. If one were not yet in the static
gauge, one would impose Neumann boundary condition for the $\rho$
direction and Dirichlet in the $u$-direction. These two, when combined
imply that string end ``orthogonally'' on the brane worldvolume ${\rm
d}u/{\rm d}\rho\rightarrow \infty$. 

Finally, the expressions for the energy and the angular momentum
carried by the string are given by
\begin{align}
\label{e:Edef}
E &= \int\!{\rm d}\sigma\, \frac{1}{\sqrt{...}} \bigg(  \left(\frac{u}{R_{\text{D4}}}\right)^3 f(u)
(\rho')^2  + u'^2 \bigg)\,, \\[1ex]
\label{e:Jdef}
J &= \int\!{\rm d}\sigma\, \frac{1}{\sqrt{...}} \omega \rho^2 \bigg(  \left( \frac{u}{R_{\text{D4}}} \right)^3\,\rho'^2 + \frac{u'^2}{f(u)} \bigg) \, ,
\end{align}
where $\sqrt{...}$ is once more the density of the Nambu-Goto action~\eqref{Sind}.

\subsection{Properties of the spectrum and melting of large-spin meson}

Due to the complexity of equation~\eqref{equ}, we will analyse the
spectrum of large spinning strings numerically. Nevertheless,
interesting qualitative features can easily be extracted from this
analysis.

\begin{figure}[t]
\begin{center}
\psfrag{rho}{$\rho$}
\psfrag{u}{$u$}
\includegraphics[width=.5\textwidth]{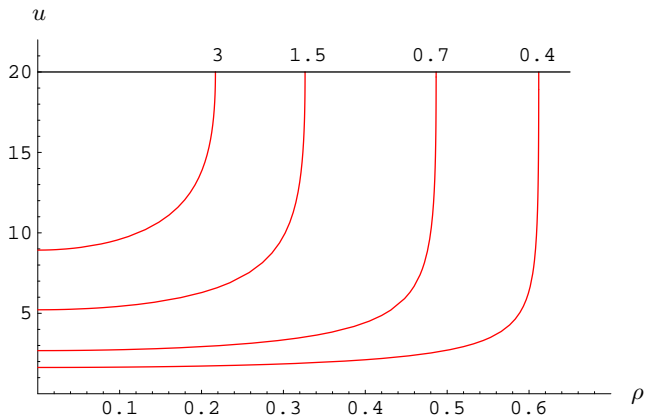}
\end{center}
\caption{Shapes of rotating strings with fixed position of the flavour
  brane~$u_0=20$ (i.e.~almost fixed $L/R$, see figure~\protect\ref{f:L_vs_yT})
  and fixed temperature~$u_T=1$, for various values of the
  frequency~$\omega$.\label{f:shapes_vs_w}}
\end{figure}

Let us first discuss the spectrum for \emph{fixed} temperature.  By
analysing the shapes of the string for various values of~$\omega$, we
see that as~$\omega$ decreases, i.e.~the spin of the meson increases,
the string endpoints get more and more separated, and the U-shaped
string penetrates deeper to the horizon. It becomes more and more
rectangular shaped. A few characteristic shapes, for mesons with
fixed~$T$ and $m_q$ but varying~$\omega$ (or equivalently~$J$) are
depicted in figure~\ref{f:shapes_vs_w}.

\begin{figure*}[t]
\begin{center}
\psfrag{E2}{$E^2$}
\psfrag{J}{$J$}
\psfrag{w}{$\omega$}
\vspace{2ex}
\includegraphics[width=.45\textwidth]{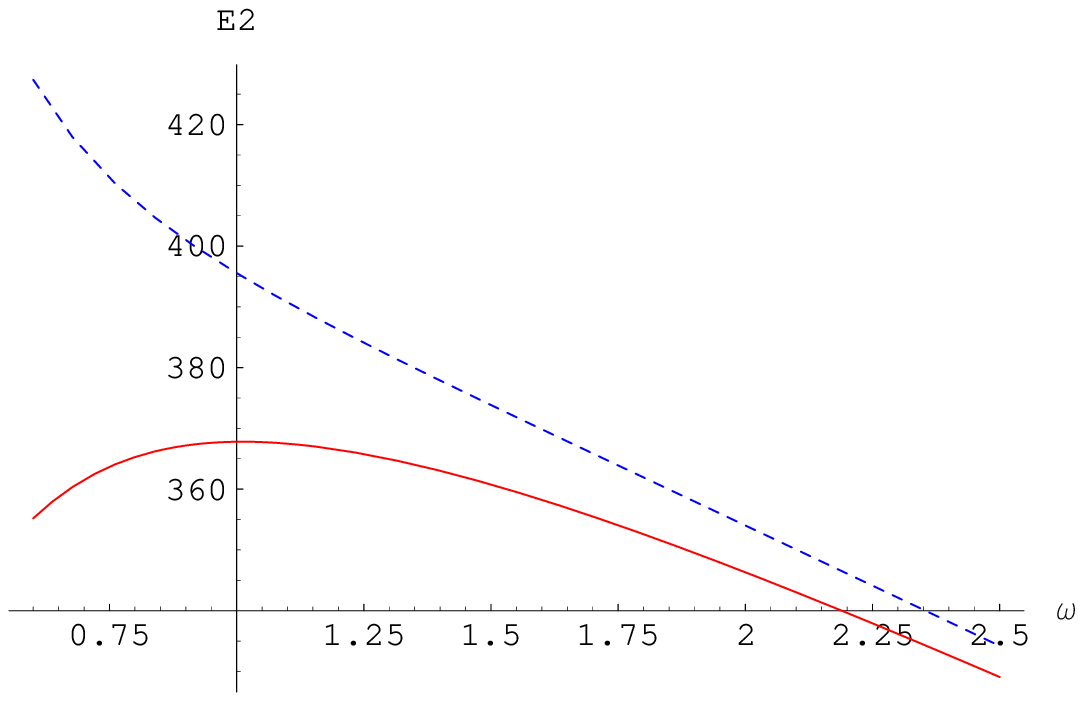}\qquad
\includegraphics[width=.45\textwidth]{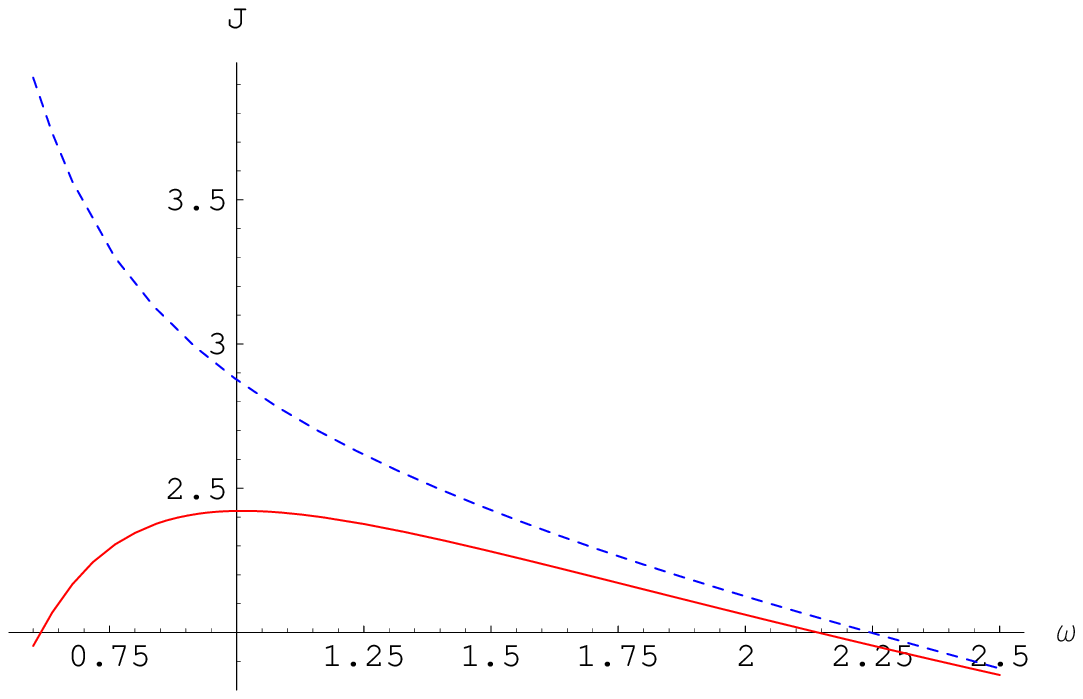}
\end{center}
\caption{The energy (squared) and angular momentum of U-shaped
strings, as a function of the frequency~$\omega$. The dashed (blue)
curves are for zero temperature, while the solid (red) curves are for
the intermediate-temperature regime.\label{f:EJom}}
\end{figure*}

Integration of the equation~\eqref{equ} yields the dispersion relation
in parametric form, i.e~$E(\omega)$ and $J(\omega)$ at fixed
temperature. These curves, together with similar curves at zero
temperature, are shown in the plots in figure~\ref{f:EJom}. We see
that at \emph{zero temperature}, the spin increases as the angular
frequency~$\omega$ is decreased (just like in flat space, where this
behaviour is essentially determined by the condition~$\omega L = 1$
which says that the endpoints move with the velocity of light). As the
spin increases, the string hangs deeper into the geometry and hence
the energy of the meson goes up.
The same happens at non-zero temperature for large values of
$\omega$, i.e.~when the string is short and thus far away from the
horizon. However, as $\omega$ is decreased, even though the length of
the string continues to grow, the effective tension of the parts of
string in the region near the horizon decreases rapidly. This effects
leads to a slower growth of spin and energy, and hence to the
appearance of a maximum in the energy and spin. This is shown in the
solid (red) curves of the plots in figure~\ref{f:EJom}. 

\begin{figure*}[t]
\begin{center}
\vspace{2ex}
\psfrag{E2}{$E^2$}
\psfrag{J}{$J$}
\psfrag{w}{$\omega$}
\includegraphics[width=.47\textwidth]{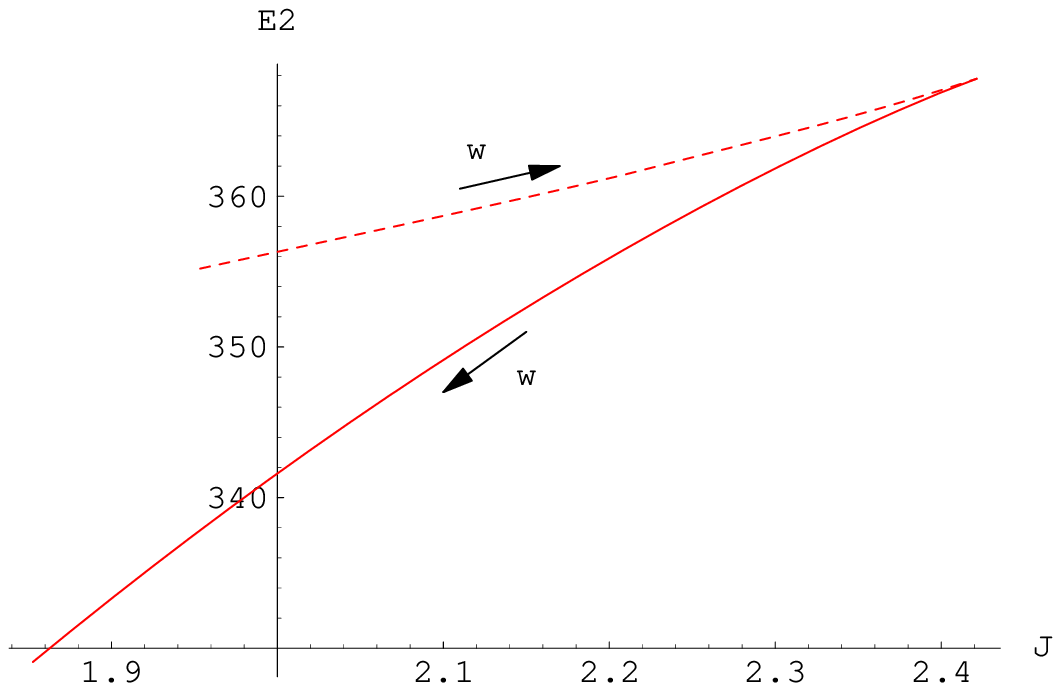}\qquad
\psfrag{w}{}
\includegraphics[width=.47\textwidth]{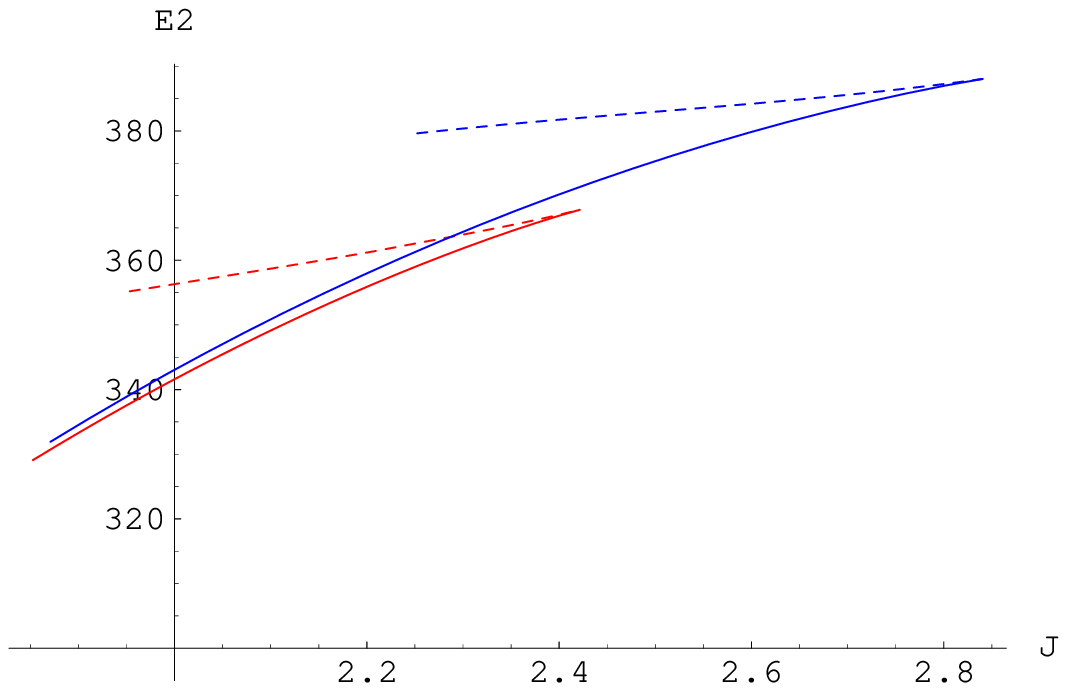}
\end{center}
\caption{Left panel: Energy versus angular momentum for $L/R=0.24$ in
  the intermediate-temperature phase,~$T/T_c=\sqrt{2}$. The lower
  segment (solid) is for $\omega>1$, the upper one (dashed) for
  $\omega<1$. There is a maximum reachable spin. Right panel: Similar
  plot, but now for two different temperatures. The blue (upper)
  curves are for a lower temperature~$T/T_c=1$, the red (lower) curves
  for a higher temperature~$T/T_c=\sqrt{2}$. Observe that for a fixed
  spin, the energies go down as the temperature goes up. \label{EJ}}
\end{figure*}

When we eliminate the variable~$\omega$ from the two plots in
figure~\ref{f:EJom}, to obtain a plot of the energy versus the spin,
we get figure~\ref{EJ}. The appearance of a maximum in energy and spin
is again clearly visible. We also see that the two states with
identical spin~$J$ actually have different energy: the ones with
smaller~$\omega$ are more energetic than the ones with
larger~$\omega$. Therefore, the upper branch to the left of the
maxima in the plots of figure~\ref{f:EJom} is presumably unstable and
will decay to the lower branch.

Apart from the two configurations in figure~\ref{EJ}, with $E_1$ and
$E_2$ for fixed $J$, there is yet another configuration one may
consider; this configuration is in spirit closer to the dissociation
of Wilson loops. Namely, one could consider the configuration of two
\emph{dragged} single quarks (to be described in more detail in the
next section), with linear velocities~$v$ which are such that the
whole system carries the same angular momentum as our mesons. Then,
one could compare the energy of such a configuration with the energy
of the meson before dissociation. However, as shown
in~\cite{Herzog:2006gh} this configuration always carries infinite
energy, and thus cannot be the one which acts as initial data for the
dissociated meson.\footnote{In this sense the dissociation of spinning
strings is different from the dissociation of Wilson
loops~\cite{Rey:1998bq,Brandhuber:1998bs}.}

The plot of $J$ versus $\omega=\omega(u_{\text{bottom}})$ is similar
to the plot of the inter-quark distance versus the minimum distance to
the horizon found for Wilson loops. For Wilson loops, it was found
that there is a maximum inter-quark distance. For spinning strings,
such a maximum distance appears too. However, this maximum distance is
not reached for the same value of~$\omega$ as the value for which the
maximum spin is attained. A priori, this is a source of worry, since
it could mean that the relation between~$J$ and the inter-quark
distance~$L$ is not monotonic. However, we have verified numerically
that the maximum inter-quark distance always occurs in the unphysical
part of the spectrum; the relation between~$J$ and~$L$ is monotonic in
the physical regime (this is somewhat similar to the monotonicity
of~$L$ versus~$y_T$ depicted in figure~\ref{f:L_vs_yT}).

\begin{figure}[t]
\begin{center}
\psfrag{Jmax}{$J_{\text{max}}$} 
\psfrag{T}{$T/T_c$} 
\psfrag{u}{$u$}
\psfrag{r}{$\rho$} 
\includegraphics[width=.9\columnwidth]{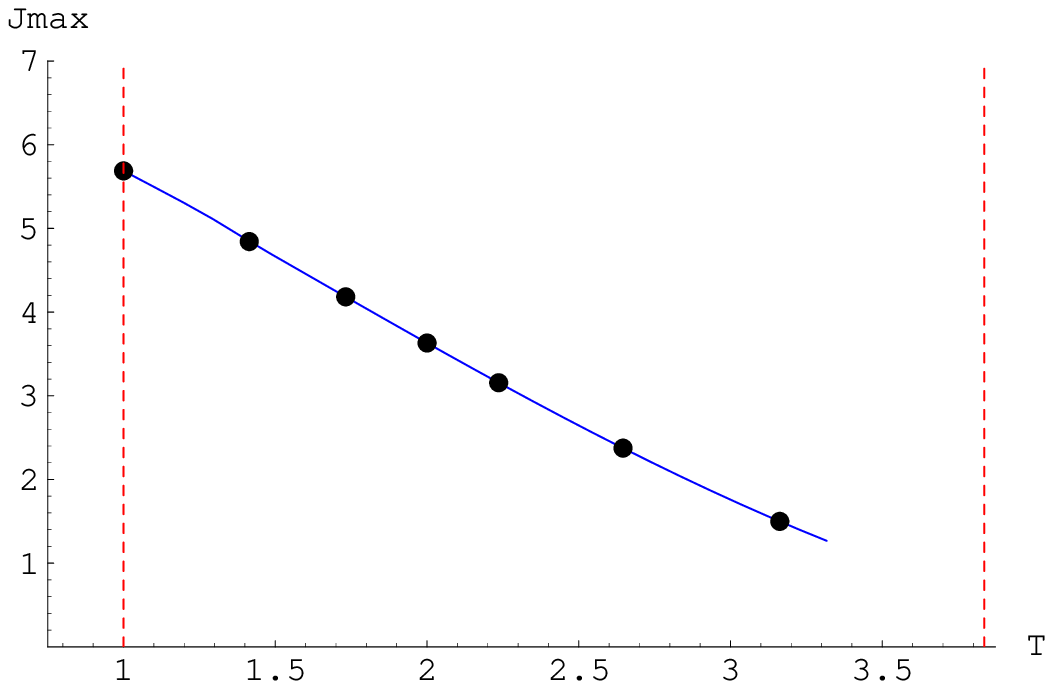}
\end{center}
\caption{Analysis of maximum spin for U-shaped
  strings in the intermediate-temperature regime, as a function of
  temperature, for fixed~$L/R=0.24$.
  Angular momentum is normalised as in~\protect\eqref{e:Jdef}. There
  is of course a divergence as~$T/T_c \rightarrow 0$, but before that
  point is reached, one encounters the phase transition to the
  low-temperature background (vertical dotted line at
  $T/T_c=1$).  \label{spinmax}}
\end{figure}

\begin{figure}[t]
\begin{center}
\psfrag{E2}{$E^2$}
\psfrag{T}{$T/T_c$}
\vspace{2ex}
\includegraphics[width=.9\columnwidth]{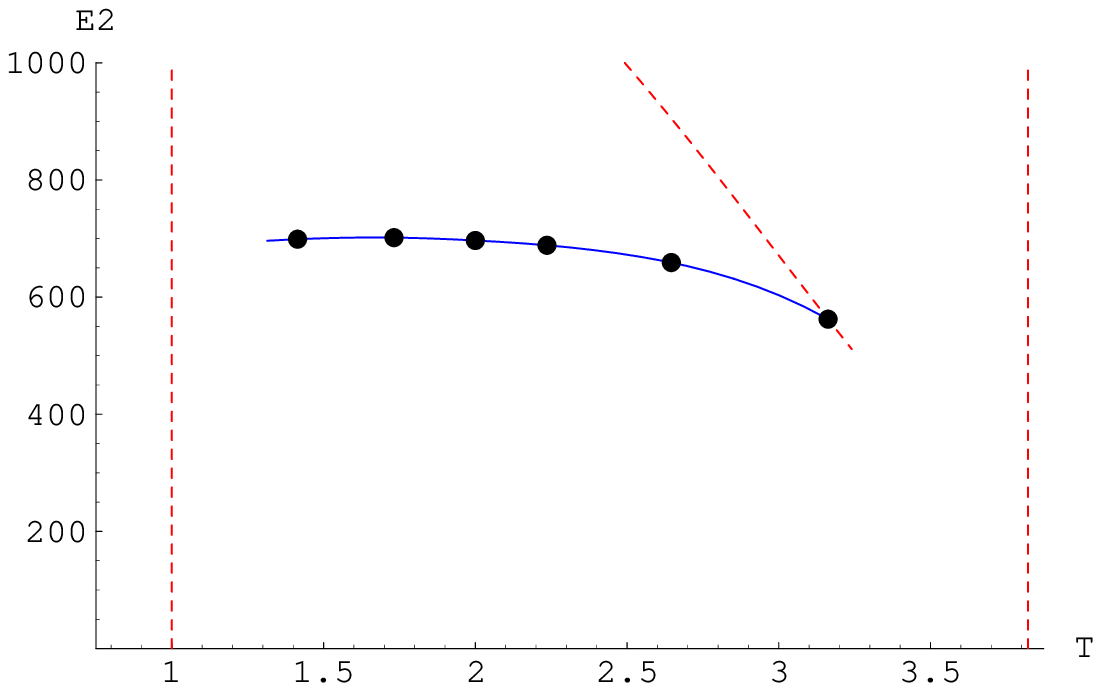}
\end{center}
\caption{Analysis of the temperature dependence of the energy of
  U-shaped strings. Displayed are the squared energies of a state of
  spin~$J\approx 1.5$, which dissociates at~$T/T_c=3.16$
  when~$L/R=0.24$. The diagonal dotted curve shows the maximum
  possible energy for a given temperature. Note the qualitative
  similarity with the low-spin plot,
  figure~\protect\ref{f:running}.\label{f:running-large}}
\end{figure}

\begin{figure*}[t]
\begin{center}
\vspace{2ex}
\psfrag{Jmax}{$J_{\text{max}}$} 
\psfrag{r}{$\rho$} 
\psfrag{u}{$u$} 
\psfrag{5.0}{$T/T_c = 2.24$} 
\psfrag{1.0}{$T/T_c = 1.00$}
\includegraphics[width=.45\textwidth]{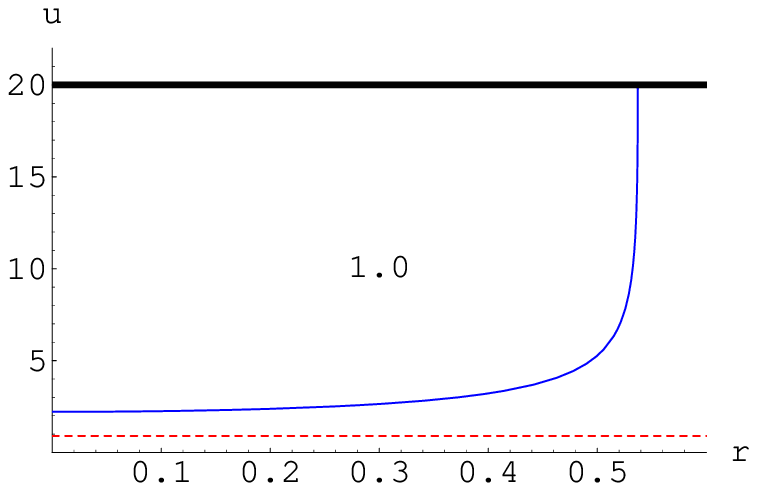}\qquad
\includegraphics[width=.45\textwidth]{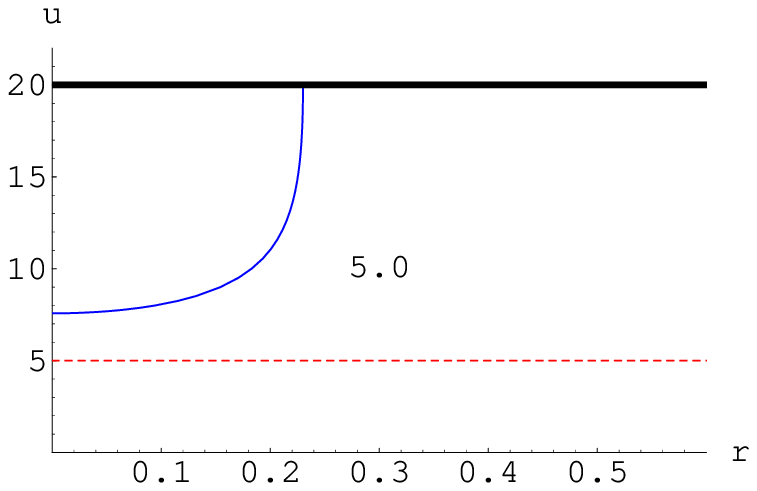}
\end{center}
\caption{The two plots illustrate the shape of the maximum-$J$
  strings at two selected temperatures (dotted horizontal red lines
  denote the horizon, solid black lines the flavour brane).\label{shapeJ}}
\end{figure*}

From the analysis above, we see that for a fixed temperature there is
a \emph{maximal} value of the spin which a meson can carry. It is
natural to interpret the temperature at which this happens as the
critical temperature at which a meson of spin~$J_{\text{max}}$ melts.
Thus we conclude that (as intuitively expected) the dissociation
temperature of large-spin mesons is \emph{spin dependent}.  As
the temperature increases, the maximal value of the spin that a meson can
carry decreases (see figure~\ref{spinmax}), i.e.~for given quark
mass, higher-spin mesons melt at lower temperature.
We also see that for meson of fixed angular momentum, the energy
decreases as a function of temperature, see
figure~\ref{f:running-large}.  This is the same behaviour as was
observed for the low-spin mesons.
\begin{figure}[t]
\begin{center}
\vspace{2ex}
\psfrag{size}{\!\!\!\!\raisebox{1ex}{\hbox{4d size}}}
\psfrag{T}{$T/T_c$}
\includegraphics[width=.9\columnwidth]{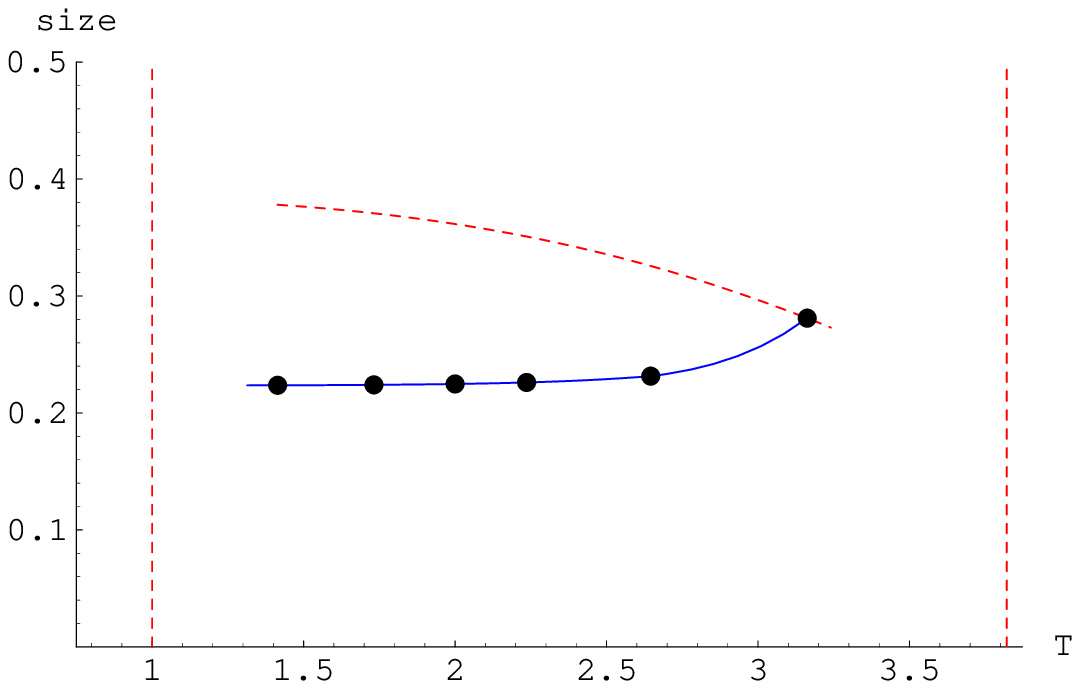}
\end{center}
\caption{The four-dimensional size of a large-spin meson of fixed
  spin as a function of $T/T_c$: when the temperature goes up, the
  meson expands. The dotted curve denotes the size of the maximum-spin
  meson at a given temperature.\label{f:size_vs_T}}
\end{figure}
However, the four-dimensional size of the meson (i.e.~the distance
between the two endpoints of the U-shaped string measured along the
probe-brane) increases as the temperature goes up; see
figure~\ref{f:size_vs_T}.  Finally, by considering the shapes of
strings at different temperatures (figure \ref{shapeJ}) we observe
that, as the temperature is increased, the largest allowed strings
become less and less rectangular shaped, and more and more U-shaped,
thus leading to the breakdown of the four-dimensional picture of a
string with two massive, point-like strings attached to
it~\cite{Kruczenski:2004me}. This is an important observation if one
wants to construct four-dimensional effective string models for mesons
at finite temperature.

\subsection{Drag effects for mesons}
\label{s:drag}

It was recently observed that the drag force experienced by free
quarks moving through a hot gluon plasma can be modelled in string
theory~\cite{Herzog:2006gh,Casalderrey-Solana:2006rq,Gubser:2006bz,Herzog:2006se,Caceres:2006dj,Friess:2006aw,Sin:2006yz}.
In this setup, the quark is a single string ending with one end on the
flavour brane and extending all the way down to the horizon. Because a
strictly vertical string moving rigidly through the background would
not have a real action~\eqref{Sind}, the string has to be bent when it
is ``pushed'' through the plasma. In addition the bent string does not
end anymore orthogonally on the brane. This means that one has to
apply a force on the string endpoint, or in other words, one has 
to ``drag'' the string in order to keep it moving.

In more detail, this conclusion follows from considering the string
action evaluated on the ``naive'' ansatz describing a rigid straight
string moving with velocity~$v_x$,
\begin{equation}
\label{ansatz-quark}
t=\tau \,, \quad u= \sigma\,, \quad x = v_x\, t \, .
\end{equation}
Plugging this ansatz into the action~\eqref{Sind}, we get
\begin{equation}
S_{\text{quark}} = \sqrt{1 - \frac{v_x^2}{f(u)}} \, .
\end{equation}
For this expression to be real, a purely vertical string cannot
stretch lower than
\begin{equation}
\label{fact1}
u > u_c  \,, \quad u_c = \frac{u_T}{(1 - v_x^2)^{1/3}} \, ,
\end{equation}
and it clearly cannot stop there when there is no second flavour brane
present. The true shape is obtained by adding a~$\sigma$-dependent
term to the ansatz for~$x$. The string now bends as it reaches down to
the horizon. The bent, moving string does not, however, satisfy
Neumann nor Dirichlet boundary conditions, neither at the horizon nor at
the probe brane. Hence the world-sheet momentum does not vanish at
the string endpoints: there is a world-sheet momentum flow into the
horizon, which has to be ``added'' by a force dragging the string on
the upper, probe brane side. This effect translates to a drag force
in the gauge theory picture.

Hence in summary, we see that there are \emph{two} effects happening
as one tries to move a single string in the hot background: firstly,
the string shape is modified in a way which depends on the temperature
and velocity, and secondly, in order to preserve the motion one needs
to apply a force.

It is a natural question to ask if  such effects are also
exerted on spinning mesons, as the temperature is turned on. One might
expect that even a simple rotating motion experiences a drag effect,
but the analysis of the previous subsection shows that this effect
does not exist. Instead, the rotating string is always sufficiently
high above the curve beyond which the action would turn imaginary. No
world-sheet momentum gets lost behind the horizon. On the other hand,
the first effect is still present. The bending of the rotating meson
does depend both on the angular velocity and on the temperature.

\begin{figure*}[t]
\begin{center}
\vspace{2ex}
\psfrag{J}{$J$}
\psfrag{omega}{$\omega$}
\psfrag{v0}{\small $v_y=0$}
\psfrag{v98}{\small $v_y=0.98$}
\psfrag{v09}{\small $v_y=0.9$}
\psfrag{rho}{$\rho$}
\psfrag{u}{$u$}
\includegraphics[width=.48\textwidth]{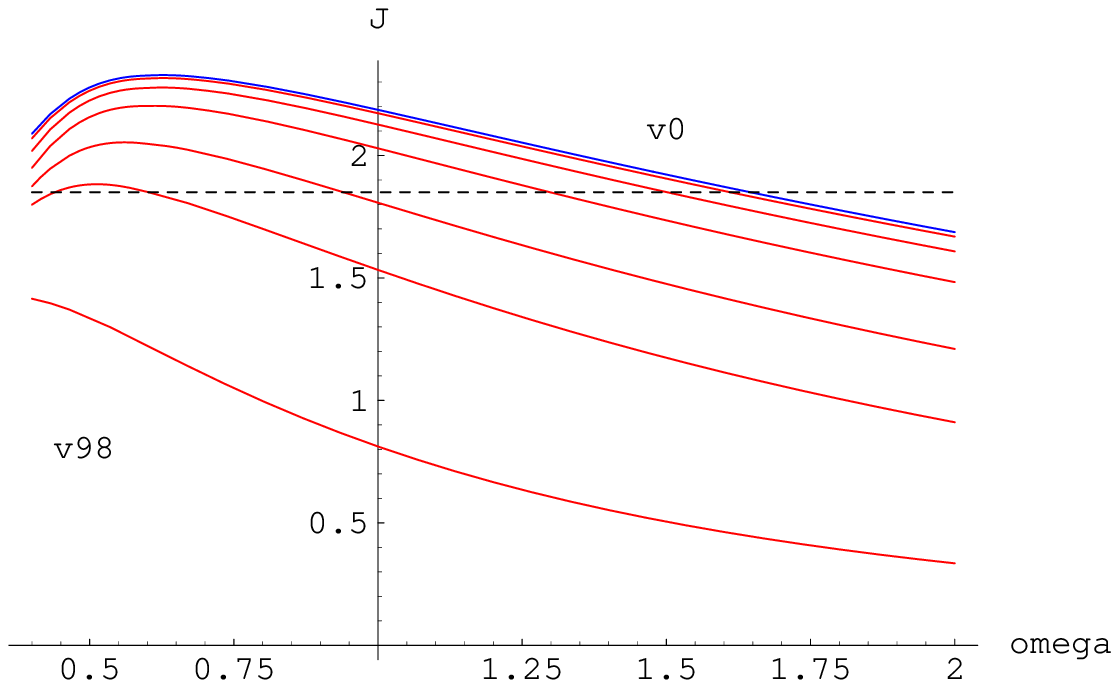}\quad
\includegraphics[width=.48\textwidth]{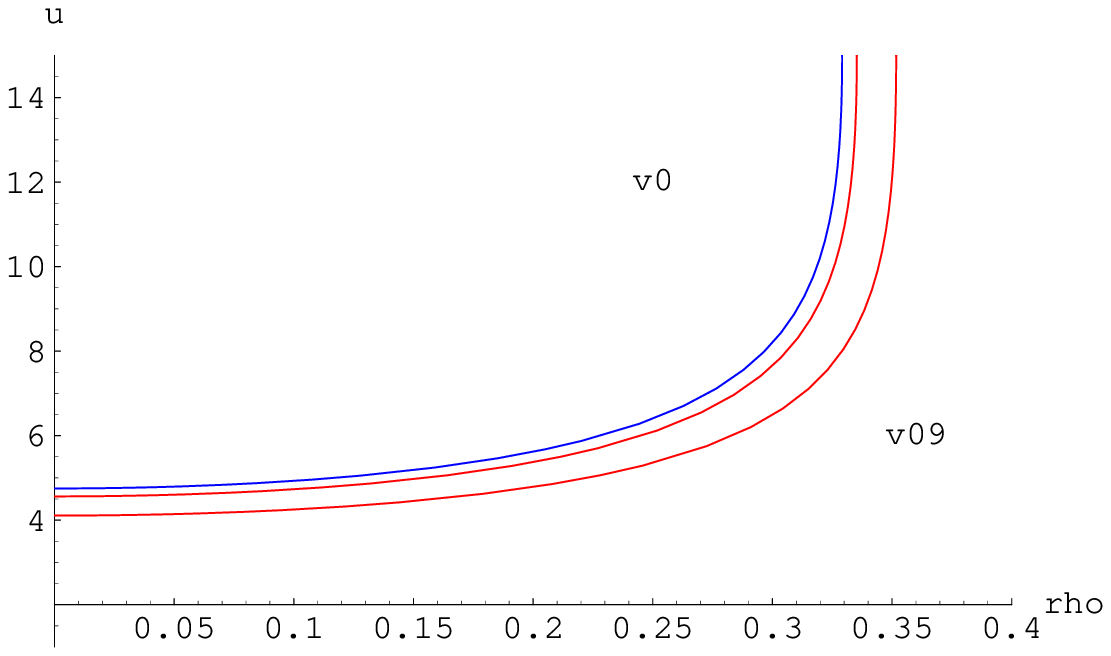}
\end{center}
\caption{Analysis of the effect of a transverse velocity on spinning
  U-shaped strings. The left panel depicts the relation between~$J$
  and~$\omega$ for fixed position of the flavour brane. The upper
  curve is for~$v_y=0$, while the other curves display results for
  increasing values~$v_y=0.2, 0.4, 0.6, 0.8, 0.9, 0.98$. The panel on
  the right shows the effect of a transverse velocity on the shape,
  keeping the quark masses and spin fixed (corresponding to the dashed
  black line in the left panel). 
\label{f:vy}}
\end{figure*}

However, there is another type of motion which one can consider, and
which is more similar to the moving-quark situation. Namely, one can
consider a linear motion of the meson in a direction orthogonal to the
plane of rotation.\footnote{One could also consider motion in the
  direction in which the string rotates. In this case the equations of
  motion are much more involved. We do, however, expect that
  the conclusion reached by studying this motion will be similar to
  those  reached by studying motion in a transverse direction.}
A suitable ansatz for this motion is
\begin{equation}
\label{e:movingmeson}
 t =\tau \, , \quad  \rho = \sigma \, , \quad u=u(\rho) \, , \quad
 \varphi  = \omega \tau  \, , 
 \quad  y = v_y\, \tau \, .
\end{equation}
In this case the action becomes
\begin{equation}
\label{Sindv}
S = \int\!{\rm d}\tau \,  {\rm d}\rho \,
\sqrt{\frac{u^3}{R^3_{\text{D4}}}\left(1  + \frac{u'^2}{f(u)}
  \frac{R_{\text{D4}}^3}{u^3}\right) \left( f(u)  - \rho^2\omega^2 - v^2\right)} \,  .
\end{equation}
The only modification with respect to the rotating meson is the
addition of a term~``$-v^2$'' to the last factor under the square
root. The condition for the action to be real is now
\begin{equation}
\label{e:minuv}
u \geq \frac{u_T}{(1-\rho^2 \omega^2 -v^2)^{1/3}}\,.
\end{equation}
Hence, if the ansatz~\eqref{e:movingmeson} makes sense, there should
exist a solution to the equations of motion such that the spinning
configuration lies entirely above this curve. If that is not the case,
the ansatz has to be modified.

We have verified that the equations of motion obtained
from~\eqref{Sindv} are the same as those of the full system after
insertion of the ansatz~\eqref{e:movingmeson}. Subsequently, we have
integrated the equations of motion using the same techniques as for
the zero-velocity case. The results are depicted in figure~\ref{f:vy}.
The left panel shows that, as the transverse speed is increased, the
maximum possible spin decreases. If we take a state of fixed spin~$J$
and look at how its shape changes as we increase the transverse speed,
we obtain the second panel in figure~\ref{f:vy} (this corresponds to
following the states which lie on the black dotted curve in the first
panel). All these configurations, up to the maximum-spin ones, lie
safely above the limiting curve~\eqref{e:minuv}. Note how the size of
a fixed-spin meson increases slightly as the transverse velocity is
turned on.

Thus, the important physical conclusion seems to be that colour
singlet states do not experience any drag effect. From the
field-theoretic point of view this makes sense, as the absence of a
colour monopole moment means that these states do not couple directly
to gluon degrees of freedom. Because of the absence of drag, mesons do
not experience any energy loss when propagating through the
quark-gluon plasma: no force is necessary to keep them moving with
fixed velocity.\footnote{We should emphasise that what we have
computed here is different from the computation
of~\cite{Sin:2006yz}. In that paper, a situation is considered in
which two quarks, being endpoints of a ``twisted'' Wilson loop, move
away from each other. The required energy influx is higher at finite
temperature, giving rise to a four-dimensional drag force. For us, the
two quarks which make up the meson move in the same direction, and we
do not change the U-shaped string as a function of time.}  Note
moreover that the moving meson does not need to bend in the direction
of motion, i.e.~there is no~$y(u)$ term present
in~\eqref{e:movingmeson}. However, the shape of the meson in
the~$(\rho,u)$ plane certainly is velocity dependent. As shown on the
right panel of figure~\ref{f:vy}, the distance to the horizon
decreases, leading to a lower melting temperature.

\begin{figure}[t]
\begin{center}
\psfrag{LatJmax}{\hspace{-2em}$L^{4d}$ at max-spin}
\psfrag{v}{$v$}
\includegraphics[width=.9\columnwidth]{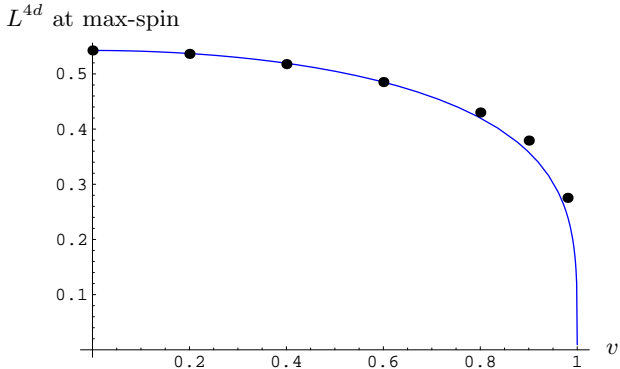}
\end{center}
\caption{The behaviour of the four-dimensional size $L^{4d}$ for maximum-spin
  mesons as a function of the transverse velocity (for a particular
  choice of~$m_q$ and~$T=T_c$). The blue curve
  depicts $L^{4d}(v=0) \cdot (1-v^2)^{1/4}$, as obtained analytically for
  Wilson loops in~\cite{Liu:2006nn}.
  \label{f:LatJmax}}
\end{figure}

Figure~\ref{f:vy} (left panel) also shows that the maximum spin goes down as the
velocity goes up. There is thus a critical velocity beyond which a
meson of fixed spin has to dissociate. Similarly, the four-dimensional
size of the meson decreases with increasing velocity, as can be seen
in figure~\ref{f:LatJmax}. The data is approximated rather well by the
relation
\begin{equation}
\label{e:LatJmax}
  L^{4d}_{\text{max-spin}}(v) \approx L^{4d}_{\text{max-spin}}(v=0)\cdot (1-v^2)^{1/4}\,.
\end{equation}
This fit was motivated by the analytic results of~\cite{Liu:2006nn},
in which a similar dependence on the velocity was found for the
screening length, or more precisely, the maximum interquark distance
for Wilson loops in an AdS black hole background. See also related
results in~\cite{Chernicoff:2006hi}.

Finally, one should keep in mind that our analysis in this section is
purely classical, and restricted to the sigma model, without any
coupling to the fields on brane. In principle there will be classical
(and quantum) energy loss due to radiation effects on brane, because
the meson is essentially a dipole. However, the interpretation of this
energy loss is different, as it corresponds to the decay of high-spin
mesons into low-spin
ones~\cite{Peeters:2005fq,Peeters:2004pt,Peeters:2005pb}, both at zero
and at non-zero temperature.

\section{Spectrum at high temperature}
\label{s:highT}

Finally, let us briefly discuss the high-temperature phase. In this phase, the
minimal free energy of the system is attained when the two stacks of
branes are disconnected (see the third panel of
figure~\ref{f:threephases}). The profile of the left and right stacks
of branes is characterised by $u'\equiv {\rm d}u/{\rm d}x^4\rightarrow
\infty$ and the induced metric on the probe branes and probe
anti-branes takes the form
\begin{multline}
\label{e:highTmetric}
{\rm d}\hat{s}^2_{\text{high}}=\left( \frac{u}{R_{\text{D4}}} \right)^{3/2}
  \left[ -f(u) {\rm d}t^2+  \delta_{ij}{\rm d}x^{i}{\rm d}x^j \right ] \\[1ex]
+\left( \frac{R_{\text{D4}}}{u} \right)^{3/2}\frac{1}{f(u)} {\rm d}u^2 
 +
 \left( \frac{R_{\text{D4}}}{u} \right)^{3/2} u^2 {\rm d}\Omega_4^2  \,,
\end{multline}
i.e.~it is the same as the background metric without the $g_{44}({\rm
d}x^4)^2$ term.  Considering now the action of the gauge fields on
each of the two stacks of branes, it is straightforward to see that
the action will be similar to the one at intermediate temperature, but
with~$\gamma=f^{-1}$. The differential equation for the modes is now
\begin{equation}
\label{e:eqhighT}
-u^{1/2}\,f(u)\,\partial_u\left( u^{5/2}
 f(u) \partial_u \psi_{(n)}\right) = R_{\text{D4}}^3 \, m_n^2\, \psi_{(n)}\,.
\end{equation}
The normalisation conditions are similarly simplified,
\begin{equation}
\label{e:normshighT}
\begin{aligned}
\int_{u_T}^\infty \!{\rm d}u\, u^4 f^{-1}(u)
\left(\frac{u}{R_{\text{D4}}}\right)^{9/2}\,\psi_{(m)}\psi_{(n)} &=
\delta_{mn}\,,\\[1ex]
\int_{u_T}^\infty \!{\rm d}u\, u^4 
\left(\frac{R_{\text{D4}}}{u}\right)^{3/2}\, \phi^{(0)} \phi^{(0)} &= 1\,.
\end{aligned}
\end{equation}

The first thing to observe now is that the massless pion disappears
from the spectrum. This mode, which would be given by~$\phi^{(0)} =
u^{-5/2} f(u)^{-1}$, is no longer normalisable. Computation of its
norm leads to the integral
\begin{equation}
\int_{u_T}^\infty\!{\rm d}u\, u^{5/2} \Big| u^{-5/2} f(u)^{-1}\Big|^2\,,
\end{equation}
which, while convergent at the upper boundary, is divergent at the
lower boundary because~$f(u) \sim \sqrt{u-u_T}$ for $u\sim u_T$. Here,
it is crucial that the D8-branes extend all the way down to the horizon
at~$u=u_T$ (recall that at finite temperature, these $f$-factors did
not lead to a divergence because the integral had a lower limit
at~$u=u_0$). In accordance with the fact that chiral symmetry is
restored in the high-temperature phase, we see that the Goldstone
boson has disappeared.

In order to see whether the remainder of the spectrum is discrete or
continuous, we write our equation~\eqref{e:eqhighT} as (we set~$R_{\text{D4}}=u_T=1$ from now)
\begin{equation}\label{spectrumhigh}
f(u)\,\partial_u\left( u^{5/2} f(u)\partial_u \psi_{(n)}\right) + u^{-1/2}\,m^2\, \psi_{(n)} = 0\,.
\end{equation}
If we expand this near~$u=1$ we get
\begin{equation}
9\,(u-1)\,\partial_u\Big( (u-1)\,\partial_u \psi(u)\Big)
  + m^2\,\psi(u) = 0\,.
\end{equation}
The general solution to this equation is
\begin{equation}
\label{e:exacthighTnh}
\psi(u) = c_1\, \cos\Big( \frac{m}{3} \log(u-1) \Big) 
        + c_2\, \sin\Big( \frac{m}{3} \log(u-1) \Big) \,.
\end{equation}
These modes are normalisabile using the norm~\eqref{e:normshighT} for
any value of the constants~$c_1$ and~$c_2$.

If one now shoots from infinity, starting with the normalisable mode
for a fixed~$m^2$, one can always match this to a regular solution at
the horizon because there are two free parameters available there.
Because there are two regular solutions at the horizon the spectrum
will be continuous in the high-temperature phase.

Note that this type of analysis could in principle be made also for
the zero- and intermediate-temperature phases. In those cases, one will
find that again both solutions are regular at the tip of the probe
brane. However, there is now an additional condition, namely that the
solution can be continued to the other side of the probe brane and
remain normalisable. So in these two phases, it is not regularity at
the tip, but rather the condition that the function is either odd or
even which makes the spectrum discrete.
\vspace{-2ex}

\section*{Note added}

After publication of the first version of this paper,
\cite{Liu:2006nn} appeared, which contains a computation of the
screening length at finite temperature (the maximum interquark
distance for Wilson loops).  Our results leading to figure~\ref{f:vy}
agree with the analytic behaviour~$L^{4d}_{\text{max-spin}}(v) \sim
  L^{4d}_{\text{max-spin}}(v=0)\, (1-v^2)^{1/4}$ found there; we have added
figure~\ref{f:LatJmax} and the paragraph around~\eqref{e:LatJmax} to
make this more explicit.
\vspace{-2ex}

\section*{Acknowledgements}

We would like to thank Ofer Aharony for many illuminating discussions
and comments on a draft of this paper, and Angel Paredes for pointing
us at a computational error. J.~S.~would like to thank E.~Shuryak
for a useful discussion.  The work of J.S.~was supported in part by
the Israel Science Foundation (grant number 03200306) and by a grant
of DIP (H.52).



\begin{thebibliography}{38}
\expandafter\ifx\csname natexlab\endcsname\relax\def\natexlab#1{#1}\fi
\expandafter\ifx\csname bibnamefont\endcsname\relax
  \def\bibnamefont#1{#1}\fi
\expandafter\ifx\csname bibfnamefont\endcsname\relax
  \def\bibfnamefont#1{#1}\fi
\expandafter\ifx\csname citenamefont\endcsname\relax
  \def\citenamefont#1{#1}\fi
\expandafter\ifx\csname url\endcsname\relax
  \def\url#1{\texttt{#1}}\fi
\expandafter\ifx\csname urlprefix\endcsname\relax\def\urlprefix{URL }\fi
\providecommand{\bibinfo}[2]{#2}
\providecommand{\eprint}[2][]{\url{#2}}

\bibitem[{\citenamefont{Sakai and Sugimoto}(2005{\natexlab{a}})}]{Sakai:2004cn}
\bibinfo{author}{\bibfnamefont{T.}~\bibnamefont{Sakai}} \bibnamefont{and}
  \bibinfo{author}{\bibfnamefont{S.}~\bibnamefont{Sugimoto}},
  \bibinfo{journal}{Prog.\ Theor.\ Phys.} \textbf{\bibinfo{volume}{113}},
  \bibinfo{pages}{843} (\bibinfo{year}{2005}{\natexlab{a}}),
  \eprint{hep-th/0412141}.

\bibitem[{\citenamefont{Sakai and Sugimoto}(2005{\natexlab{b}})}]{Sakai:2005yt}
\bibinfo{author}{\bibfnamefont{T.}~\bibnamefont{Sakai}} \bibnamefont{and}
  \bibinfo{author}{\bibfnamefont{S.}~\bibnamefont{Sugimoto}}
  (\bibinfo{year}{2005}{\natexlab{b}}), \eprint{hep-th/0507073}.

\bibitem[{\citenamefont{Aharony et~al.}(2006)\citenamefont{Aharony,
  Sonnenschein, and Yankielowicz}}]{Aharony:2006da}
\bibinfo{author}{\bibfnamefont{O.}~\bibnamefont{Aharony}},
  \bibinfo{author}{\bibfnamefont{J.}~\bibnamefont{Sonnenschein}},
  \bibnamefont{and}
  \bibinfo{author}{\bibfnamefont{S.}~\bibnamefont{Yankielowicz}}
  (\bibinfo{year}{2006}), \eprint{hep-th/0604161}.

\bibitem[{\citenamefont{Parnachev and Sahakyan}(2006)}]{Parnachev:2006dn}
\bibinfo{author}{\bibfnamefont{A.}~\bibnamefont{Parnachev}} \bibnamefont{and}
  \bibinfo{author}{\bibfnamefont{D.~A.} \bibnamefont{Sahakyan}}
  (\bibinfo{year}{2006}), \eprint{hep-th/0604173}.

\bibitem[{\citenamefont{de~Forcrand et~al.}(2001)}]{deForcrand:2000jx}
\bibinfo{author}{\bibfnamefont{P.}~\bibnamefont{de~Forcrand}}
  \bibnamefont{et~al.} (\bibinfo{collaboration}{QCD-TARO}),
  \bibinfo{journal}{Phys.\ Rev.} \textbf{\bibinfo{volume}{D63}},
  \bibinfo{pages}{054501} (\bibinfo{year}{2001}), \eprint{hep-lat/0008005}.

\bibitem[{\citenamefont{Karsch et~al.}(2006)\citenamefont{Karsch, Kharzeev, and
  Satz}}]{Karsch:2005nk}
\bibinfo{author}{\bibfnamefont{F.}~\bibnamefont{Karsch}},
  \bibinfo{author}{\bibfnamefont{D.}~\bibnamefont{Kharzeev}}, \bibnamefont{and}
  \bibinfo{author}{\bibfnamefont{H.}~\bibnamefont{Satz}},
  \bibinfo{journal}{Phys.\ Lett.} \textbf{\bibinfo{volume}{B637}},
  \bibinfo{pages}{75} (\bibinfo{year}{2006}), \eprint{hep-ph/0512239}.

\bibitem[{\citenamefont{Wong}(2006)}]{Wong:2006dz}
\bibinfo{author}{\bibfnamefont{C.-Y.} \bibnamefont{Wong}}
  (\bibinfo{year}{2006}), \eprint{hep-ph/0606200}.

\bibitem[{\citenamefont{Karsch}(2000)}]{Karsch:1999vy}
\bibinfo{author}{\bibfnamefont{F.}~\bibnamefont{Karsch}},
  \bibinfo{journal}{Nucl.\ Phys.\ Proc.\ Suppl.} \textbf{\bibinfo{volume}{83}},
  \bibinfo{pages}{14} (\bibinfo{year}{2000}), \eprint{hep-lat/9909006}.

\bibitem[{\citenamefont{Shuryak}(2004)}]{Shuryak:2003xe}
\bibinfo{author}{\bibfnamefont{E.}~\bibnamefont{Shuryak}},
  \bibinfo{journal}{Prog.\ Part.\ Nucl.\ Phys.} \textbf{\bibinfo{volume}{53}},
  \bibinfo{pages}{273} (\bibinfo{year}{2004}), \eprint{hep-ph/0312227}.

\bibitem[{\citenamefont{Kalinovsky et~al.}(2005)\citenamefont{Kalinovsky,
  Radzhabov, and Volkov}}]{Kalinovsky:2005kx}
\bibinfo{author}{\bibfnamefont{Y.~L.} \bibnamefont{Kalinovsky}},
  \bibinfo{author}{\bibfnamefont{A.~E.} \bibnamefont{Radzhabov}},
  \bibnamefont{and} \bibinfo{author}{\bibfnamefont{M.~K.} \bibnamefont{Volkov}}
  (\bibinfo{year}{2005}), \eprint{hep-ph/0508264}.

\bibitem[{\citenamefont{Herzog et~al.}(2006)\citenamefont{Herzog, Karch,
  Kovtun, Kozcaz, and Yaffe}}]{Herzog:2006gh}
\bibinfo{author}{\bibfnamefont{C.~P.} \bibnamefont{Herzog}},
  \bibinfo{author}{\bibfnamefont{A.}~\bibnamefont{Karch}},
  \bibinfo{author}{\bibfnamefont{P.}~\bibnamefont{Kovtun}},
  \bibinfo{author}{\bibfnamefont{C.}~\bibnamefont{Kozcaz}}, \bibnamefont{and}
  \bibinfo{author}{\bibfnamefont{L.~G.} \bibnamefont{Yaffe}}
  (\bibinfo{year}{2006}), \eprint{hep-th/0605158}.

\bibitem[{\citenamefont{Casalderrey-Solana and
  Teaney}(2006)}]{Casalderrey-Solana:2006rq}
\bibinfo{author}{\bibfnamefont{J.}~\bibnamefont{Casalderrey-Solana}}
  \bibnamefont{and} \bibinfo{author}{\bibfnamefont{D.}~\bibnamefont{Teaney}}
  (\bibinfo{year}{2006}), \eprint{hep-ph/0605199}.

\bibitem[{\citenamefont{Gubser}(2006)}]{Gubser:2006bz}
\bibinfo{author}{\bibfnamefont{S.~S.} \bibnamefont{Gubser}}
  (\bibinfo{year}{2006}), \eprint{hep-th/0605182}.

\bibitem[{\citenamefont{Herzog}(2006)}]{Herzog:2006se}
\bibinfo{author}{\bibfnamefont{C.~P.} \bibnamefont{Herzog}}
  (\bibinfo{year}{2006}), \eprint{hep-th/0605191}.

\bibitem[{\citenamefont{Caceres and Guijosa}(2006)}]{Caceres:2006dj}
\bibinfo{author}{\bibfnamefont{E.}~\bibnamefont{Caceres}} \bibnamefont{and}
  \bibinfo{author}{\bibfnamefont{A.}~\bibnamefont{Guijosa}}
  (\bibinfo{year}{2006}), \eprint{hep-th/0605235}.

\bibitem[{\citenamefont{Friess et~al.}(2006)\citenamefont{Friess, Gubser, and
  Michalogiorgakis}}]{Friess:2006aw}
\bibinfo{author}{\bibfnamefont{J.~J.} \bibnamefont{Friess}},
  \bibinfo{author}{\bibfnamefont{S.~S.} \bibnamefont{Gubser}},
  \bibnamefont{and}
  \bibinfo{author}{\bibfnamefont{G.}~\bibnamefont{Michalogiorgakis}}
  (\bibinfo{year}{2006}), \eprint{hep-th/0605292}.

\bibitem[{\citenamefont{Sin and Zahed}(2006)}]{Sin:2006yz}
\bibinfo{author}{\bibfnamefont{S.-J.} \bibnamefont{Sin}} \bibnamefont{and}
  \bibinfo{author}{\bibfnamefont{I.}~\bibnamefont{Zahed}}
  (\bibinfo{year}{2006}), \eprint{hep-ph/0606049}.

\bibitem[{\citenamefont{Neri and Gocksch}(1983)}]{Neri:1983ic}
\bibinfo{author}{\bibfnamefont{F.}~\bibnamefont{Neri}} \bibnamefont{and}
  \bibinfo{author}{\bibfnamefont{A.}~\bibnamefont{Gocksch}},
  \bibinfo{journal}{Phys.\ Rev.} \textbf{\bibinfo{volume}{D28}},
  \bibinfo{pages}{3147} (\bibinfo{year}{1983}).

\bibitem[{\citenamefont{Pisarski}(1984)}]{Pisarski:1983db}
\bibinfo{author}{\bibfnamefont{R.~D.} \bibnamefont{Pisarski}},
  \bibinfo{journal}{Phys.\ Rev.} \textbf{\bibinfo{volume}{D29}},
  \bibinfo{pages}{1222} (\bibinfo{year}{1984}).

\bibitem[{\citenamefont{Schreiber}(2004)}]{Schreiber:2004ie}
\bibinfo{author}{\bibfnamefont{E.}~\bibnamefont{Schreiber}}
  (\bibinfo{year}{2004}), \eprint{hep-th/0403226}.

\bibitem[{\citenamefont{Shifman}(2005)}]{Shifman:2005zn}
\bibinfo{author}{\bibfnamefont{M.}~\bibnamefont{Shifman}}
  (\bibinfo{year}{2005}), \eprint{hep-ph/0507246}.

\bibitem[{\citenamefont{Karch et~al.}(2006)\citenamefont{Karch, Katz, Son, and
  Stephanov}}]{Karch:2006pv}
\bibinfo{author}{\bibfnamefont{A.}~\bibnamefont{Karch}},
  \bibinfo{author}{\bibfnamefont{E.}~\bibnamefont{Katz}},
  \bibinfo{author}{\bibfnamefont{D.~T.} \bibnamefont{Son}}, \bibnamefont{and}
  \bibinfo{author}{\bibfnamefont{M.~A.} \bibnamefont{Stephanov}}
  (\bibinfo{year}{2006}), \eprint{hep-ph/0602229}.

\bibitem[{\citenamefont{Gottlieb et~al.}(1997)}]{Gottlieb:1996ae}
\bibinfo{author}{\bibfnamefont{S.~A.} \bibnamefont{Gottlieb}}
  \bibnamefont{et~al.}, \bibinfo{journal}{Phys.\ Rev.}
  \textbf{\bibinfo{volume}{D55}}, \bibinfo{pages}{6852} (\bibinfo{year}{1997}),
  \eprint{hep-lat/9612020}.

\bibitem[{\citenamefont{Ghoroku and Yahiro}(2005)}]{Ghoroku:2005kg}
\bibinfo{author}{\bibfnamefont{K.}~\bibnamefont{Ghoroku}} \bibnamefont{and}
  \bibinfo{author}{\bibfnamefont{M.}~\bibnamefont{Yahiro}}
  (\bibinfo{year}{2005}), \eprint{hep-ph/0512289}.

\bibitem[{\citenamefont{Bhattacharyya and Raha}(1995)}]{Bhattacharyya:1995xt}
\bibinfo{author}{\bibfnamefont{A.}~\bibnamefont{Bhattacharyya}}
  \bibnamefont{and} \bibinfo{author}{\bibfnamefont{S.}~\bibnamefont{Raha}},
  \bibinfo{journal}{J.\ Phys.} \textbf{\bibinfo{volume}{G21}},
  \bibinfo{pages}{741} (\bibinfo{year}{1995}).

\bibitem[{\citenamefont{Asakawa and Hatsuda}(2004)}]{Asakawa:2003re}
\bibinfo{author}{\bibfnamefont{M.}~\bibnamefont{Asakawa}} \bibnamefont{and}
  \bibinfo{author}{\bibfnamefont{T.}~\bibnamefont{Hatsuda}},
  \bibinfo{journal}{Phys.\ Rev.\ Lett.} \textbf{\bibinfo{volume}{92}},
  \bibinfo{pages}{012001} (\bibinfo{year}{2004}), \eprint{hep-lat/0308034}.

\bibitem[{\citenamefont{Datta et~al.}(2004)\citenamefont{Datta, Karsch,
  Petreczky, and Wetzorke}}]{Datta:2003ww}
\bibinfo{author}{\bibfnamefont{S.}~\bibnamefont{Datta}},
  \bibinfo{author}{\bibfnamefont{F.}~\bibnamefont{Karsch}},
  \bibinfo{author}{\bibfnamefont{P.}~\bibnamefont{Petreczky}},
  \bibnamefont{and} \bibinfo{author}{\bibfnamefont{I.}~\bibnamefont{Wetzorke}},
  \bibinfo{journal}{Phys.\ Rev.} \textbf{\bibinfo{volume}{D69}},
  \bibinfo{pages}{094507} (\bibinfo{year}{2004}), \eprint{hep-lat/0312037}.

\bibitem[{\citenamefont{Wissel et~al.}(2006)\citenamefont{Wissel, Laermann,
  Shcheredin, Datta, and Karsch}}]{Wissel:2005pb}
\bibinfo{author}{\bibfnamefont{S.}~\bibnamefont{Wissel}},
  \bibinfo{author}{\bibfnamefont{E.}~\bibnamefont{Laermann}},
  \bibinfo{author}{\bibfnamefont{S.}~\bibnamefont{Shcheredin}},
  \bibinfo{author}{\bibfnamefont{S.}~\bibnamefont{Datta}}, \bibnamefont{and}
  \bibinfo{author}{\bibfnamefont{F.}~\bibnamefont{Karsch}},
  \bibinfo{journal}{PoS} \textbf{\bibinfo{volume}{LAT2005}},
  \bibinfo{pages}{164} (\bibinfo{year}{2006}), \eprint{hep-lat/0510031}.

\bibitem[{\citenamefont{Iida et~al.}(2006)\citenamefont{Iida, Doi, Ishii, and
  Suganuma}}]{Iida:2005ea}
\bibinfo{author}{\bibfnamefont{H.}~\bibnamefont{Iida}},
  \bibinfo{author}{\bibfnamefont{T.}~\bibnamefont{Doi}},
  \bibinfo{author}{\bibfnamefont{N.}~\bibnamefont{Ishii}}, \bibnamefont{and}
  \bibinfo{author}{\bibfnamefont{H.}~\bibnamefont{Suganuma}},
  \bibinfo{journal}{PoS} \textbf{\bibinfo{volume}{LAT2005}},
  \bibinfo{pages}{184} (\bibinfo{year}{2006}), \eprint{hep-lat/0509129}.

\bibitem[{\citenamefont{Morrin~et al.}(2006)}]{Morrin:2005zq}
\bibinfo{author}{\bibfnamefont{R.}~\bibnamefont{Morrin~et al.}},
  \bibinfo{journal}{PoS} \textbf{\bibinfo{volume}{LAT2005}},
  \bibinfo{pages}{176} (\bibinfo{year}{2006}), \eprint{hep-lat/0509115}.

\bibitem[{\citenamefont{Kruczenski et~al.}(2005)\citenamefont{Kruczenski,
  Zayas, Sonnenschein, and Vaman}}]{Kruczenski:2004me}
\bibinfo{author}{\bibfnamefont{M.}~\bibnamefont{Kruczenski}},
  \bibinfo{author}{\bibfnamefont{L.~A.~P.} \bibnamefont{Zayas}},
  \bibinfo{author}{\bibfnamefont{J.}~\bibnamefont{Sonnenschein}},
  \bibnamefont{and} \bibinfo{author}{\bibfnamefont{D.}~\bibnamefont{Vaman}},
  \bibinfo{journal}{JHEP\,} \textbf{\bibinfo{volume}{06}}, \bibinfo{pages}{046}
  (\bibinfo{year}{2005}), \eprint{hep-th/0410035}.

\bibitem[{\citenamefont{Rey et~al.}(1998)\citenamefont{Rey, Theisen, and
  Yee}}]{Rey:1998bq}
\bibinfo{author}{\bibfnamefont{S.-J.} \bibnamefont{Rey}},
  \bibinfo{author}{\bibfnamefont{S.}~\bibnamefont{Theisen}}, \bibnamefont{and}
  \bibinfo{author}{\bibfnamefont{J.-T.} \bibnamefont{Yee}},
  \bibinfo{journal}{Nucl.\ Phys.} \textbf{\bibinfo{volume}{B527}},
  \bibinfo{pages}{171} (\bibinfo{year}{1998}), \eprint{hep-th/9803135}.

\bibitem[{\citenamefont{Brandhuber et~al.}(1998)\citenamefont{Brandhuber,
  Itzhaki, Sonnenschein, and Yankielowicz}}]{Brandhuber:1998bs}
\bibinfo{author}{\bibfnamefont{A.}~\bibnamefont{Brandhuber}},
  \bibinfo{author}{\bibfnamefont{N.}~\bibnamefont{Itzhaki}},
  \bibinfo{author}{\bibfnamefont{J.}~\bibnamefont{Sonnenschein}},
  \bibnamefont{and}
  \bibinfo{author}{\bibfnamefont{S.}~\bibnamefont{Yankielowicz}},
  \bibinfo{journal}{Phys.\ Lett.} \textbf{\bibinfo{volume}{B434}},
  \bibinfo{pages}{36} (\bibinfo{year}{1998}), \eprint{hep-th/9803137}.

\bibitem[{\citenamefont{Liu et~al.}(2006)\citenamefont{Liu, Rajagopal, and
  Wiedemann}}]{Liu:2006nn}
\bibinfo{author}{\bibfnamefont{H.}~\bibnamefont{Liu}},
  \bibinfo{author}{\bibfnamefont{K.}~\bibnamefont{Rajagopal}},
  \bibnamefont{and} \bibinfo{author}{\bibfnamefont{U.~A.}
  \bibnamefont{Wiedemann}} (\bibinfo{year}{2006}), \eprint{hep-ph/0607062}.

\bibitem[{\citenamefont{Chernicoff et~al.}(2006)\citenamefont{Chernicoff,
  Garcia, and Guijosa}}]{Chernicoff:2006hi}
\bibinfo{author}{\bibfnamefont{M.}~\bibnamefont{Chernicoff}},
  \bibinfo{author}{\bibfnamefont{J.~A.} \bibnamefont{Garcia}},
  \bibnamefont{and} \bibinfo{author}{\bibfnamefont{A.}~\bibnamefont{Guijosa}}
  (\bibinfo{year}{2006}), \eprint{hep-th/0607089}.

\bibitem[{\citenamefont{Peeters et~al.}(2006)\citenamefont{Peeters,
  Sonnenschein, and Zamaklar}}]{Peeters:2005fq}
\bibinfo{author}{\bibfnamefont{K.}~\bibnamefont{Peeters}},
  \bibinfo{author}{\bibfnamefont{J.}~\bibnamefont{Sonnenschein}},
  \bibnamefont{and} \bibinfo{author}{\bibfnamefont{M.}~\bibnamefont{Zamaklar}},
  \bibinfo{journal}{JHEP\,} \textbf{\bibinfo{volume}{02}}, \bibinfo{pages}{009}
  (\bibinfo{year}{2006}), \eprint{hep-th/0511044}.

\bibitem[{\citenamefont{Peeters et~al.}(2004)\citenamefont{Peeters, Plefka, and
  Zamaklar}}]{Peeters:2004pt}
\bibinfo{author}{\bibfnamefont{K.}~\bibnamefont{Peeters}},
  \bibinfo{author}{\bibfnamefont{J.}~\bibnamefont{Plefka}}, \bibnamefont{and}
  \bibinfo{author}{\bibfnamefont{M.}~\bibnamefont{Zamaklar}},
  \bibinfo{journal}{JHEP\,} \textbf{\bibinfo{volume}{11}}, \bibinfo{pages}{054}
  (\bibinfo{year}{2004}), \eprint{hep-th/0410275}.

\bibitem[{\citenamefont{Peeters et~al.}(2005)\citenamefont{Peeters, Plefka, and
  Zamaklar}}]{Peeters:2005pb}
\bibinfo{author}{\bibfnamefont{K.}~\bibnamefont{Peeters}},
  \bibinfo{author}{\bibfnamefont{J.}~\bibnamefont{Plefka}}, \bibnamefont{and}
  \bibinfo{author}{\bibfnamefont{M.}~\bibnamefont{Zamaklar}},
  \bibinfo{journal}{Found.\ Phys.} \textbf{\bibinfo{volume}{53}},
  \bibinfo{pages}{640} (\bibinfo{year}{2005}), \eprint{hep-th/0501165}.

\end{thebibliography}

\end{document}